\documentclass[12pt,preprint]{aastex}

\usepackage{chngcntr}

\usepackage{color}
\usepackage{textcomp}
\usepackage[normalem]{ulem}
\usepackage{rotating}
\usepackage{lscape}

\shortauthors{Zelaya {\it et~al.\/}}
\shorttitle{Continuum Polarization of Type Ia SNe}

\begin{document}

\title{Continuum Foreground Polarization and Na~I Absorption in Type Ia SNe\footnote{Based on observations made with ESO Telescopes at the Paranal Observatory under programs 068.D-0571(A), 069.D-0438(A), 070.D-0111(A), 076.D-0178(A), 079.D-0090(A), 080.D-0108(A), 081.D-0558(A), 085.D-0731(A) and 086.D-0262(A). Also based on observations collected at the German-Spanish Astronomical Center, Calar Alto (Spain).}}
\author{
P. Zelaya\altaffilmark{2,3}, 
A. Clocchiatti\altaffilmark{3,2}, 
D. Baade\altaffilmark{4}, 
P. H{\"o}flich\altaffilmark{5}, 
J. Maund\altaffilmark{6},
F. Patat\altaffilmark{4}, 
J.R. Quinn\altaffilmark{3}, 
E.Reilly\altaffilmark{6},
L. Wang\altaffilmark{7},
J.C. Wheeler\altaffilmark{8}, 
F. F{\"o}rster\altaffilmark{2,9}, and
S. Gonz\'alez-Gait\'an\altaffilmark{2,9}
}

\email{pazelaya@astro.puc.cl}

\altaffiltext{2}{MAS - Millennium Institute of Astrophysics, Casilla 36-D,7591245, Santiago, Chile}

\altaffiltext{3}{Instituto de Astrof\'isica, Pontificia Universidad Cat\'olica de Chile, Casilla 306, Santiago 22, Chile}

\altaffiltext{4}{ESO - European Organization for Astronomical Research in the Southern Hemisphere, Karl-Schwarzschild-Str.2, 85748 Garching b. M\"unchen, Germany}

\altaffiltext{5}{Department of Physics, Florida State University, Tallahassee, Florida 32306-4350,USA}

\altaffiltext{6}{Department of Physics and Astronomy, University of Sheffield, Hicks Building, Hounsfield Road, Sheffield, S3 7RH, UK}

\altaffiltext{7}{Department of Physics, Texas A\&M University, College Station, Texas 77843-4242, USA}

\altaffiltext{8}{Department of Astronomy and McDonald Observatory, The University of Texas at Austin, 1 University Station, C1400, Austin, Texas 78712, USA}

\altaffiltext{9}{Departamento de Astronom\'ia, Universidad de Chile, Casilla 36-D, Santiago, Chile}

\begin{abstract}
We present a study of the continuum polarization over the 400--600~nm range
of 19 Type Ia SNe obtained with FORS at the VLT.
We separate them in those that show Na~I~D lines
at the velocity of their hosts and those that do not.
Continuum polarization of the sodium sample near maximum light displays a broad range of values,
from extremely polarized cases like SN~2006X to almost unpolarized ones like SN~2011ae.
The non--sodium sample shows, typically, smaller polarization values.
The continuum polarization of the sodium sample in the 400--600~nm range is linear with wavelength and 
can be characterized by the mean polarization (P$_{\rm{mean}}$). Its values span a wide range and show a linear correlation
with color, color excess, and extinction in the visual band.
Larger dispersion correlations were found with the equivalent width of the Na~I~D and Ca~II H \& K lines, 
and also a noisy relation between P$_{\rm{mean}}$ and $R_{V}$,
the ratio of total to selective extinction.
Redder SNe show stronger continuum polarization, with larger color excesses and extinctions.
We also confirm that high continuum polarization is associated with small values of $R_{V}$. 

The correlation between extinction and polarization -- and polarization angles --
suggest that the dominant fraction of dust polarization is imprinted 
in interstellar regions of the host galaxies.

We show that Na~I~D lines from foreground matter in the SN host are usually associated with non-galactic
ISM, challenging the typical assumptions in foreground interstellar polarization models.

\end{abstract}

\section{Introduction}

Since the proposal of \citet{Wheeler1971Ap&SS..11..373W}, Type Ia supernovae have been
firmly established as the result of the thermonuclear runaway of a carbon oxygen (CO) white dwarf
(WD) that has accreted matter from a donor star until reaching the
Chandrasekhar mass limit.
Significant features of the progenitor system and the explosion, however, remain unsolved.
Questions like what the companion star is, how fast the mass is accreted, at
what mass does the CO WD explode, 
how is the explosion triggered, and how does it burn its nuclear fuel, 
are still subject to debate. 
There are two popular progenitor models:
the single degenerate (SD) model \citep{Whelan1973ApJ...186.1007W,Nomoto1982ApJ...253..798N,Iben1996ApJS..105..145I},
where a WD accretes mass from a non degenerate star,
and the double degenerate (DD) model \citep{Webbink1984ApJ...277..355W,Iben1984ApJS...54..335I,Pakmor2012ApJ...747L..10P},
where two CO WDs violently merge.
Both SD and DD scenarios have collected favorable, albeit at times contradictory, observational
evidence.

Since the SD and DD models imply a different circumstellar environment, observations with the potential of
diagnosing the local neighborhood of the SNe have the potential of distinguishing between the two
progenitor channels.
The presence of variable components of the Na~I~D narrow absorption lines in some SNe, like SN~1999cl, 2006X
and 2007le \citep[respectively]{{Blondin2009ApJ...693..207B},Patat et2009A&A...508..229P,Simon2009ApJ...702.1157S},
the statistics of velocity shifts of these lines \citep{Sternberg2011Sci...333..856S,Maguire2013MNRAS.436..222M},
studies of the circumstellar matter (CSM) in supernova remnants like Kepler
\citep{Burkey2013ApJ...764...63B},
and
the blue light excess found in SN~2012cg by \citet{Marion2015arXiv150707261M}
, have been linked to the SD scenario all suggesting that the presence of a rich CSM is associated with
an evolving, non--degenerate, companion star.
On the other hand, the extremely near SN~2011fe, which was first observed just a couple of
hours after explosion, puts strong constraints on the luminosity and mass of the companion.
A normal non degenerate star is practically ruled out in favor of the DD scenario
\citep{Li2011Natur.480..348L,Pakmor2012ApJ...747L..10P}. 

Finally, the whole connection of variable Na~I~D lines with the SD scenario has
been brought into question by \citet{Soker2014}.
He proposes that the variability of the Na~I~D lines results from gas being
adsorbed by, or deadsorbed from, dust grains.

Spectropolarimetry allows for new insights into the problem of the progenitor systems
because it provides information on the SNe and the material in their lines of sight
that is not available in the usual intensity versus wavelength observations.
As routine spectropolarimetry of Type Ia SNe goes through its second decade,
it is fairly safe to say that, within the constraints of the signal-to-noise ratio,
the observation of appreciable continuum polarization in normal events is rare.
Although there have been some detections, most of them can be associated with peculiar SNe
or comfortably explained as the result of typical foreground interstellar polarization.
\citet{Howell2001ApJ...556..302H} found a peak of 0.8\% intrinsic continuum polarization for SN 1999by, higher 
than the $< 0.4$\% expected for normal Type Ia's \citep{Wang_and_Wheeler2008ARA&A}.
For this subluminous event, the polarization peaks in the red side of the spectrum and decreases to the blue, 
as predicted by the model of \citet{Wang1997ApJ...476L..27W}.
Additionally \citet{Patat2012A&A...545A...7P} found intrinsic continuum polarization $\sim 0.7\%$
for SN~2005ke, also growing from blue to red wavelengths.
SN~2005hk showed more typical lower levels of intrinsic continuum polarization, 0.36\% and 0.18\%
reported by \citet{Chornock2006PASP..118..722C} and \citet{Maund2010ApJ...722.1162M}, respectively.
These three SNe were classified as peculiar, SN 1999by and SN~2005ke are subluminous events and SN~2005hk corresponds to 2002cx-like SNe \citep{Li2003PASP..115..453L}.
From the study of a sample of four SNe, \citet{Leonard2005ApJ...632..450L}
suggested that normal, and perhaps overluminous, SNe displayed weak continuum
polarization ($\lesssim 0.4 \%$) and that subluminous ones showed larger values, although still 
modest ($\sim 0.8\%$).

SN~2006X is particularly noteworthy in this context.
\citet{Patat et2009A&A...508..229P} found that it had a strong continuum polarization,
which rose sharply towards the blue and remained constant over the 49 days covered by
the observations.
The atypical dependence with wavelength suggested an unusual origin for the polarization,
but the constancy in time indicated that it was not intrinsic to the SN.
The wavelength dependence could not be successfully matched by a Serkowski law \citep{Serkowski1975ApJ...196..261S} that incorporates the relation between wavelength of maximum polarization,
$\lambda_{\rm{max}}$, 
and a \textquotedblleft constant\textquotedblright $K$ found by
\citet{Whittet1992ApJ...386..562W}, which is generally used to describe the polarization
of the Milky Way interstellar medium (ISM).
In spite of this, the polarization was reasonably well aligned with a spiral arm of the host at
the projected position of the SN.
Polarization of light transmitted through the ISM is known to align
parallel to the magnetic field lines \citep[e.g.][]{Mathewson_Ford_1970MmRAS..74..139M}.
This is explained as the effect of extinction by elongated dust grains aligned with the
interstellar magnetic field by the paramagnetic relaxation mechanism of
\citet{Davis1951ApJ...114..206D}, which could be greatly accelerated by radiative torques
and possibly slowed down by active formation of hydrogen molecules on the grain surface
\citep{Draine_Weingartner_1997_ApJ_480_633}.
Magnetic fields, in addition, are known to align with large scale structures of galaxies, like spiral arms \citep[e.g.][]{Mathewson_Ford_1970MmRAS..74..139M,Scarrott_et_al_1987MNRAS.224..299S}.
Accordingly, \citet{Patat et2009A&A...508..229P} interpreted the continuum polarization
of SN~2006X as foreground polarization of interstellar origin, even though it had a very peculiar
wavelength dependence.
They suggested that this peculiarity resulted from abnormal dust, and related  it to the peculiar 
reddening laws found for many Type Ia SNe \citep{Wang2006CBET..396....2W,WangX2009ApJ...699L.139W,Folatelli2010AJ....139..120F}.
In addition, high resolution spectroscopy for this SN showed that the bulk of reddening came from a cold molecular cloud, 
which strengthens the idea of unusual dust in that particular region. 
\citet{Patat et2009A&A...508..229P} found that the steep rise to the blue of the polarization spectrum was reasonably well
matched by a Serkowski law with a very small $\lambda_{\rm{max}}$ and $K \sim 1.1$, a pair of
values that does not satisfy the relation of \citet{Whittet1992ApJ...386..562W}. 
Because of these odd parameters, \citet{Patat et2009A&A...508..229P} chose to model the continuum polarization
with a cubic function to represent the interstellar polarization (ISP).
The procedure is sensible but highlights how difficult it is to uncover any
continuum polarization intrinsic to the SN when the foreground polarization is atypical.
How sensitive the estimates of intrinsic polarization are with respect to the choice of ISP
is illustrated by \citet[][see their Fig.~10]{Leonard2000ApJ...536..239L}.

SNe 2008fp and 2014J were not as highly polarized as SN~2006X but showed a similar continuum polarization spectrum.
\citet{Patat2015A&A...577A..53P} combined the spectropolarimetic data of these three events with the
multicolor polarimetry of SN~1986G, and performed a comparative analysis trying to understand the properties
of extragalactic dust in the foreground of heavily reddened Type~Ia SNe.
They fitted their polarization spectrum with a Serkowski law and compared their $K$
\textquotedblleft constant\textquotedblright
and $\lambda_{\rm{max}}$ with those of Galactic stars.
They found that the $K,\lambda_{\rm{max}}$ pairs display an atypical interstellar behavior, and connected the small
$\lambda_{\rm{max}}$ values with the presence of smaller than normal dust grains.
Although the existence of circumstellar matter could not be excluded for these events, 
\citet{Patat2015A&A...577A..53P} concluded that the dust responsible for the bulk of the polarization
is interstellar.
The reason why the highly extincted SNe show a polarization spectrum so different from those of extincted stars in
the Galaxy was left open.

\citet{Hoang2015arXiv151001822H} tries to answer whether pure ISM or combined CSM plus ISM models are a better
fit to the observations of these SNe.
He uses a theoretical model of the dust grains with the size distribution and alignment as parameters.
He confirms that the low $R_{V}$ values are a consequence of an enlarged proportion of small dust grains,
and finds that SNe 2006X and 1986G are better matched by pure foreground ISM while 2014J should
carry a CSM like component.
His results are not completely conclusive in the case of SN 2008fp.

In this work, we combine the previously mentioned spectropolarimetric observations of SN~2006X, SN~2008fp and SN~2014J
with those of another 16 SNe observed with the VLT close to maximum light and at least one later epoch, to study the
continuum polarization of suspected foreground origin in SNe with both large and small reddening.
The high S/N ratio required to measure polarization allowed us to obtain high quality low resolution intensity spectra
and made it possible to detect unresolved narrow lines of IS origin in some of them.
We present the polarimetric spectra of these 19 SNe and show that 12 of them display unresolved Na~I~D absorption lines
at wavelengths consistent with the redshift of their hosts.
We do a closer analysis of the continuum polarization of these 12 SNe together with the multicolor broadband polarimetry
of SN~1986G.

In Section \ref{obs_results} we briefly introduce the observations, which will be described individually in future works, 
as well as additional data that are relevant to this paper.
In Section \ref{results} we present our results concerning the polarization spectrum, reddening, and
foreground absorption lines, in
Section \ref{discuss} we discuss them and in Section \ref{conclusions} we summarize and present our conclusions.

\section{Observations and ancillary data}\label{obs_results}

We introduce here the polarimetric spectra of the 19 Type~Ia SNe.
Eighteen of them were observed with the VLT at ESO\textasciiacute s Paranal Observatory and the remainder, SN~2014J, with the 2.2m Telescope at Calar
Alto Observatory.

\setcounter{footnote}{9}

SN~2014J, was observed with the Calar Alto Faint Object Spectrograph (CAFOS)\footnote{http://w3.caha.es/CAHA/Instruments/CAFOS/}.
A preliminary analysis of these data was done by \citet{Patat_et_al_2014a} and presented as well in
\citet{Patat2015A&A...577A..53P}. A full description of this observation will be given elsewhere.

The VLT observations were performed with two different spectrographs.
Before April 2008 we used the FOcal Reducer and low dispersion
Spectrograph, FORS1, at the UT2 Kueyen 8.2m Telescope \citep{Appenzeller1998Msngr..94....1A}, and after that date,
we used FORS2, which inherited the polarimeter on FORS1 and was mounted at the UT1 Antu 8.2m Telescope.
All the observations were performed employing the same four half-wave plate angles (0, 22.5, 45 and 67.5 degrees), a 1.1 arc-sec slit and the low resolution (112 \AA{}/mm) grism G300V without an order separation filter attached, with the exception of SN~2006X and 2008fp for which we used the GG435 filter.
A standard reduction procedure was followed (bias subtraction, flat fielding and wavelength calibration) with the help of IRAF\footnote{ http://iraf.noao.edu---IRAF is distributed by the National Optical Astronomy Observatories,which are operated by the Association of Universities for Research in Astronomy, Inc., under cooperative agreement with the National Science Foundation.},
and sensitivity curves were computed for each supernova from flux standards observed at the same
epoch at a position angle of the retarder plate of 0 degrees. Flux calibration and Stokes parameters calculations were
performed employing our own IDL routines following the procedure described by \citet{PatatRomaniello2006PASP..118..146P} and the ESO FORS manual\footnote{For details of the instrument check the FORS Manual at\\ http://www.eso.org/sci/facilities/paranal/instruments/fors/doc/}. 
The resolution of the spectra, measured from the width of sky emission lines at $\sim5900$~\AA\ is 13.4~\AA.

At least one subsequent observation, from $\sim$7 to $\sim$49 days after maximum
was obtained for all of SNe in Table~\ref{tab_obslog}.
These later observations and a more comprehensive study of those introduced here, will be
provided in forth-coming publications.

We found that twelve of the 19 SNe of our sample displayed unresolved
Na~I~D lines at velocities consistent with those of the parent galaxies.
This subset of SNe is called the \textquotedblleft sodium sample\textquotedblright ~in what follows.
Some details of the observations, the SNe, and the parent galaxies, are presented in Table~\ref{tab_obslog},
including the equivalent width (EW) of the narrow, unresolved, Na~I~D lines.
These were calculated using the SPLOT task in IRAF.
The seven SNe for which we did not find foreground Na~I~D lines at the redshift of the host
will be called the \textquotedblleft non-sodium sample\textquotedblright ~in what follows.
We found that 11 out of 12 SNe in our sodium sample displayed narrow interstellar Ca~II H \& K lines as well\footnote{
When our spectra did not reach the wavelength range of the calcium lines we used those of the public
SN database of the Center for Astrophysics \citep{Blondin2012AJ....143..126B} and those of the Carnegie Supernova Project (CSP)
\citep{Folatelli2013ApJ...773...53F}.}.
We measure the EW of these lines using the script described in \citet{Forster2012ApJ...754L..21F} and the results
are also reported in Table~\ref{tab_obslog}.

Multicolor photometry is available for eleven of the twelve SNe in the sodium sample.
\citet{Phillips2013ApJ...779...38P} studied a sample of 32 Type~Ia SNe with multiband photometry, 
in some cases extending to the infrared, and high resolution spectroscopy
of the foreground Na~I~D and K~I~$\lambda \lambda 7655, 7699$ lines.
They provide estimates of color excess and $R_{V}$, and hence $A_{V}$,
for six of the 12.
\citet{Burns2014} present in detail the calculation of these parameters.
Six of our events, SN~2002fk, SN~2003W, SN~2002bo, SN~2014J, SN~2005hk and SN~2011ae,
were not included in \citet{Phillips2013ApJ...779...38P}.
Upon our request, Chris Burns has kindly calculated
$E(B-V)$, $R_{V}$, and $A_{V}$ values and/or upper limits
for the first two (SN~2002fk and SN~2003W).
For the next two, SN~2002bo and SN~2014J, we rely on the values provided by \citet{Amanullah2014} and
\citet{Krisciunas2004AJ....128.3034K}, and for SN~2005hk we follow \citet{Phillips2007PASP..119..360P}.
For SN~2011ae there were no public photometric data.
We estimated its color at maximum from our first spectrum taken two days after maximum.
The spectrum was corrected for the redshift of the parent galaxy and reddening in our Galaxy \citep{Schlegel1998ApJ...500..525S}.
We then computed a synthetic color by numerical integration through the $B$ and $V$ pass-bands
given by \citet{Landolt1992AJ....104..372L}.
The color uncertainty was obtained by integrating the spectrum flux with its
reported observational error (see Table~\ref{tab_phot}).
For this SN we were unable to find extinction measurements.

Finally, we incorporated SN~1986G which exploded in Centaurus~A
and was intensively observed. A wealth of data is publicly available, including multicolor broad-band polarimetry.
The data relevant to this paper are presented in Appendix~\ref{ap:86G}.

\section{Results and Analysis}\label{results}

\begin{figure}[t!]
\centering
\includegraphics[width=13cm]{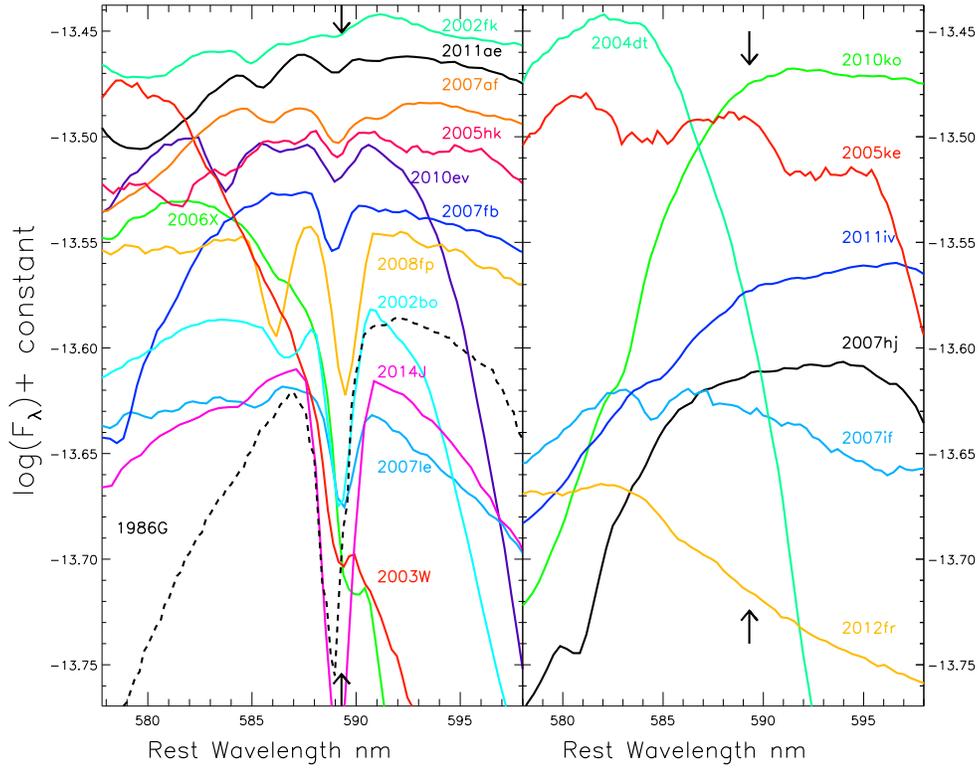}
\caption{ Rest-frame low-resolution intensity spectra of the 19 SNe in our sample.
The left panel shows the sodium sample, and the right panel the no-sodium one.
The position of the Na~D~I lines at the redshift of the SN hosts is indicated by arrows.
}
\label{fig:Nalines}
\end{figure}

The intensity spectra of the 19 SNe, in a rest frame wavelength range bracketing the Na~I~D lines,
are displayed in Figure~\ref{fig:Nalines}.
The left panel shows the sodium sample and the right one the non-sodium sample.
The logarithmic vertical scale of both panels is the same.
The left panel displays a large diversity of EWs, from very small as in the case of SN~2003W,
up to very large as in the cases of SN~2002bo or SN~2014J.
Some SNe display, in addition, Na~I~D lines produced by interstellar matter in our Milky Way.
In the case of SN~2014J, the Na~I~D lines of the Milky Way and those of the parent galaxy appear blended
at the resolution of our observations.
High resolution spectra, however, reveal that the Galactic component is negligible in comparison with 
that of the host \citep{Sternberg2014, Patat2015A&A...577A..53P}.

The right panel reveals that SNe in the non-sodium sample show no trace of absorption at the expected
rest frame wavelength of the Na~I~D lines.
To quantify the absence with upper limits we fitted gaussian profiles of the spectra resolution at the
expected wavelengths of the lines.
We found that the EW of the Na~I~D line is smaller than 0.03~\AA\ for SN~2007if and smaller than 0.01~\AA\ 
for the other six SNe in the non-sodium sample.
Some of the SNe in the non-sodium sample display Na~I~D lines of matter in the Milky Way.
Since, in the sodium sample, host galaxy Na~I~D lines are typically at least as strong as the galactic ones,
this suggests that the non-detections are real.

Figures \ref{fig:fig1} and \ref{fig:fig1b} 
exhibit the polarization and flux spectra of the SNe in the sodium and non-sodium samples, respectively.
The observations in both figures illustrate the variety of spectropolarimetric properties of Type~Ia SNe
near maximum light.
Some of them exhibit strong line polarization while some others display a relatively
featureless polarization spectrum.
The analysis of the line polarization is out of the scope of the present work, where we study the continuum polarization spectra.

We decided to constrain the analysis of the polarization spectra to the wavelength region between 420 and 580~nm.
We did so firstly because it is a region with good signal-to-noise ratio in our spectra, secondly
because the continuum polarization of our sodium sample tends to be higher in this range, and finally,
because most of the imaging polarimetry in our Galaxy and neighbors has been observed in the $V$
band passband, which makes this a convenient interval to facilitate comparisons. 
This wavelength region of interest has been shaded in grey in Figures \ref{fig:fig1} and \ref{fig:fig1b}.
The area of the shaded regions, which is proportional to the average polarization in the range,
was used to organize the SNe in a sequence of decreasing average polarization from the
upper left to the lower right panels.

Comparing the gray areas of the continuum polarization spectra of SNe in Figure
\ref{fig:fig1} with those of Figure \ref{fig:fig1b} reveals that most of the SNe in the sodium sample show continuum polarization with a negative slope,
and that this slope decreases together with the area of the shaded region.
Also, for most SNe in the sodium sample, the shape of the polarization spectrum is approximately linear
in the selected range. This is shown by a fitted cyan line (to be defined below).
These characteristics seem to be independent of the presence of line polarization, which for
some SNe is obvious (e.g. SN~2007le and 2002bo in the sodium sample, and 2004dt and 2007hj in the 
non-sodium one) while for some others is not so clear (e.g. SN~2008fp in the sodium sample and
2007if in the non-sodium one).
Finally, there is some continuity between Figs.~\ref{fig:fig1} and \ref{fig:fig1b}.
The continuum polarization spectra of SNe with small average polarization (i.e. those in the lower panels
of Fig.~\ref{fig:fig1}), tend to be more similar to the continuum polarization spectra in Fig.~\ref{fig:fig1b},
while those of SNe with sizable mean polarization (i.e. those in the upper panels
of Fig.~\ref{fig:fig1}) are quite different.

All SNe in the sodium sample have later-time spectropolarimetric observations,
extending from about a week up to several weeks after maximum light (seven in the case of SN~2006X).
These later observations
are partially shown in Figure~\ref{fig:EvolPmeanNaD}, where the observed continuum polarization of each SN is computed as the average polarization in the region between 400 and 600~nm, with rejection of 3-$\sigma$ deviant values.
The figure shows that, irrespective of the evolution of the SN light curves and flux spectrum, the polarization varies by less than 0.05\% on average.
The exception is SN~2011ae, which shows a variation of 0.08\% in the continuum polarization between our two epochs.
We take this variation as an evidence of intrinsic continuum polarization and, since the expectations are that this polarization will decrease as the SN ages \citep{Wang_and_Wheeler2008ARA&A}, we
estimated the ISP using the latest observation.
This is why both the Serkowski and the Serkowski-Whittet profiles in Fig.~\ref{fig:fig1} fall below the continuum polarization of the first epoch.

Another important observable of the polarimetric signal is the direction of the polarization angle.
In the wavelength region between 420 and 580~nm the direction of polarization is wavelength independent.
There is no rotation of the polarization angle.

Also, the polarization pseudo-vectors of almost the entire sodium sample are aligned with major features of the hosts 
projected at the position of the SNe, like the spiral arms or the disc of the host galaxy.
This is particularly relevant in the case of SN~2008fp, which shows a prominent Na~I~D line caused from matter in
the Milky Way.
As noted before by \citet{Cox2014A&A...565A..61C}, 
a contribution to the ISP from our galaxy is expected, but the orientation of the observed polarization
suggests that it is not very significant.
Orientation with host structures are not so evident in the non-sodium sample
where the polarization angles seem to be randomly oriented with
respect to major features of the hosts.
This is due to the low polarization values in that sample, which are associated with large error bars
(see Figures~\ref{fig:Host_Na} and ~\ref{fig:Host_NoNa}), and also to the relatively larger impact of
Galactic polarization components. 

Indications are, then, that the observed continuum polarization of the SNe in the sodium sample is ISP,
albeit of a peculiar character, produced predominantly by foreground matter in the host galaxy.
Even though the ISP of some SNe might have a Galactic component, we fit a unique ISP and assume it belongs mainly 
to the host galaxy. 
The assumption is supported by both, the alignment of the polarization angle with the local structure of the host (see Fig.~\ref{fig:Host_Na}),
the Na~I absorption lines of extragalactic origin which are always dominant compared to the Galactic ones (see Fig.~\ref{fig:Nalines}),
and is further justified in Section \ref{discuss}.

Accordingly, we attempted to fit these polarization spectra with 
a single \citet{Serkowski1975ApJ...196..261S} interstellar polarization law:
\begin{equation} \label{eq:Serkowski}
 P(\lambda) = P_{\rm{max}}\,\exp \left( -K\,\ln^2 \frac{\lambda_{\rm{max}}}{\lambda} \right),
\end{equation}

\noindent
where $P_{\rm{max}}$, the maximum polarization, and $\lambda_{\rm{max}}$, the wavelength at
which the maximum polarization occurs, are free parameters, and $K$ is a constant.
We take $K = 1.15$ from \citet{Serkowski1975ApJ...196..261S}.
These curves are shown by a dashed blue line in Fig.~\ref{fig:fig1}.
The dot-dashed green lines correspond to the modification of the Serkowski profile proposed
by \citet{Whittet1992ApJ...386..562W}, in which the constant $K$ is replaced by
\begin{equation} \label{eq:Serkowski-Whittet}
K=0.01 \pm 0.05 + (1.66 \pm 0.09)\lambda_{\rm{max}},
\end{equation}

\noindent with $\lambda_{\rm{max}}$ in microns.

Fitting either of Equations \ref{eq:Serkowski} or \ref{eq:Serkowski-Whittet} 
to our observations requires
values of $\lambda_{\rm{max}}$ in the range $0.20-0.40$~$\mu$m for eight of the
12 SNe in the sodium sample.
This wavelength range is unusually short in comparison with the typical values found for the ISP
of the Milky Way ($\lambda_{\rm{max}} \sim 0.55$~$\mu$m).
Three of the remaining, SNe~2011ae, 2002fk and 2005hk, have Milky Way-like values of the peak wavelength,
and SN~2007af demands a longer wavelength ($\lambda_{\rm{max}} \simeq 0.75~\mu$m).

The $\lambda_{\rm max}$ parameters of the SNe in the sodium sample make them an atypical group when compared 
with the standards of the Milky Way.
This is better illustrated by the histograms shown in Figure~\ref{fi:lambdamax-histogram} where we compare the
distribution of $\lambda_{\rm max}$ for the sodium and non-sodium samples with that of stars in the Galaxy
\citep[from][]{Serkowski1975ApJ...196..261S}.
Finally, Fig.~\ref{fig:fig1} shows that the Serkowski-Whittet polarization spectrum gives
a poor fit to the observed polarization, especially in those objects that have
higher polarization in the blue continuum.
This had been quantitatively shown by \citet{Patat2015A&A...577A..53P} for the highly polarized SNe 2006X, 2008fp
and 2014J (see their Figure~2).

It is seen in Fig.~\ref{fig:fig1} that, for the small values of $\lambda_{\rm{max}}$ found,
the original Serkowski law approaches a straight line, especially in the blue part of the
spectrum.
We have found it most useful to build upon this similarity, and describe the observed
polarization by modeling it with a straight line fitted to the wavelength range
between 420 nm and 580 nm (the region shaded in gray in Figs.~\ref{fig:fig1} and \ref{fig:fig1b}).
Line blanketing in this range is strong and even if the scattering photosphere was asymmetric,
and its emission was intrinsically polarized,
the blanketing by different lines in different regions of the overlying expanding atmosphere,
mainly by iron lines, will provide for almost complete depolarization
\citep{Howell2001ApJ...556..302H,Wang1997ApJ...476L..27W}.
Hence, the observed continuum polarization in this region can be reasonably attributed to foreground sources.
Finally, characterizing the foreground polarization in this region facilitates the comparison with observations 
made with imaging polarimeters, typically performed at B and/or V bands.

Irrespective of the previous considerations on the continuum polarization, some of our SNe do show line polarization
in the wavelength range between 420~nm and 580~nm.
This is particularly clear in SNe with smaller overall polarization,
like SN~2004dt and 2007hj in the non-sodium sample.
The line polarization is typically associated with the absorption throughs of Si~III~4560, Si~II~5051, the S~II multiplets
between 5468 and 5654 \AA\ and, in some cases, the Mg~II line at 4481 \AA.
To reduce the influence of this line polarization in our estimates of the continuum polarization we performed a two step
iteration of the linear fit, excluding in the second step the points that were $2\sigma$ above the fit of the first step.

The linear fits of polarization in our chosen wavelength range, can be written as
\begin{equation} \label{eq:straightfit}
 P_{\rm{cont}}(\lambda)= a + b\lambda, \\ \\
\end{equation}
and are shown in Fig.~\ref{fig:fig1} and ~\ref{fig:fig1b} by a cyan line.
We choose to characterize the observations by the slope $b$ of the fitted
straight lines and the value of the mean continuum polarization $P_{\rm{mean}}$,
\begin{equation}
P_{\rm{mean}}= P_{\rm{cont}}(500\,\rm{nm}),
\end{equation}
\noindent
since 500 nm is the average wavelength of the selected region.

We understand the observed continuum polarization as the result of anisotropic extinction by aligned dust grains.
If so, the slope $b$ in the selected wavelength range depends on the physical processes that polarize the light
both at microscopic and macroscopic scales.
The intrinsic properties of the ensemble of dust grains, like size and asphericity, determine
the relative polarizing efficiencies at different wavelengths, and, in particular, in the $[420,580]$nm wavelength range.
If $\lambda_{\rm max}$ is to the blue of 420nm, then $b < 0$, and
the closer $\lambda_{\rm max}$ is to $\sim500$nm the shallower the slope $b$ will be.
On the other hand, for $\lambda_{\rm max}$ not within the $[420,580]$nm wavelength range,
a larger efficiency of grain alignment, a more ordered magnetic field, or larger amounts of dust, imply a larger $P_{\rm{max}}$
and, hence, a steeper slope $b$
(more negative if $\lambda_{\rm max} \lesssim 420$nm and more positive if $\lambda_{\rm max} \gtrsim 580$nm).
For $\lambda_{\rm max}$ close to, or within, the $[420,580]$nm wavelength range and increase of
$P_{\rm{max}}$ will not result in an appreciable change of the slope $b$.

The mean continuum polarization, $P_{\rm{mean}}$, is a measure of the average polarization
over a broad wavelength range, akin to a broad pass-band filter measurement.
For small values of polarization it should be an additive quantity, in the Stokes parameters' space,
summing over the contributions of different polarizing regions.
If the light goes through different dusty regions with magnetic fields well organized over large distance scales,
the size of $P_{\rm{mean}}$ will be a combined measure of dust column density and alignment efficiency.
If the dust properties of different regions are similar, and result in similar values of $\lambda_{\rm max}$, the combined
polarization will have a larger $P_{\rm{max}}$ and steeper slope for wavelengths smaller enough, or larger enough, than $\lambda_{\rm max}$.
If the dust properties of different regions are different, and result in different values of $\lambda_{\rm max}$, the combined
polarization will have a larger $P_{\rm{max}}$, as well.
But there will be a broader maximum in the polarization spectrum and the slope at wavelengths bluer or redder than the maximum will
be shallower.
Finally, the combined contribution of regions with randomly oriented magnetic fields should 
statistically result in no net polarization, although the combination of few regions with different dust properties
could result in rotation of the polarization angle with wavelength.

We found that the $b$ and $P_{\rm{mean}}$ parameters are well correlated (see Figure~\ref{fig:fig2}).
The correlation coefficient of the linear regression is $-0.88$ when all the objects with spectropolarimetric
observations are used\footnote{SN~1986G is excluded from all computations of correlation coefficients, because it
was observed with a very different technique.}. The regression is shown with a solid line and indicates a tight fit.
Both SNe 1986G and 2006X appear to be outliers, showing a polarization larger than expected
from the slope of the blue continuum.
If SN~2006X is removed from the sample (we stress that SN~1986G is displayed in the figures, but not included
in the fits), the correlation coefficient becomes $-0.90$ (shown by a dashed line in the figure).
Either with or without SN~2006X the probability of obtaining these correlation coefficients by chance if
$b$ and $P_{\rm{mean}}$ were not correlated is smaller than one percent.
The implied linear relation allows us to use either $b$ or $P_{\rm{mean}}$
as a single parameter to describe the observed continuum polarization.
We choose to use $P_{\rm{mean}}$ in what follows because it is easier to interpret.

Having $P_{\rm{mean}}$ as the single parameter to describe the amount of observed polarization,
we can look for
correlations between it and other observables in Tables~\ref{tab_obslog} and \ref{tab_phot}.
We have found that $P_{\rm{mean}}$ correlates with the observed color at maximum
(see Fig.~\ref{fig:Pmean_BVmax}).
The correlation coefficient for a linear regression when all the SNe are used in the fit is $0.90$,
and when SN~2006X is excluded is $0.95$.
We have also found that $P_{\rm{mean}}$ correlates very well with the extinction and color excess
measured by \citet{Phillips2013ApJ...779...38P} and \citet{Burns2014} (see Fig.~\ref{fig:Av_Pmean} and
\ref{fig:E_BVvsPmean}).
The correlation coefficients for a linear regression between $P_{\rm{mean}}$ and extinction 
when all the SNe are fitted is $0.89$, and increases to $0.91$ when SN~2006X is excluded.
For the linear fit between $P_{\rm{mean}}$ and $E(B-V)$ the correlation coefficients result $0.82$ and $0.85$ for the cases including and excluding SN~2006X, respectively.
The probability of obtaining these correlation coefficients by chance if the observables were not correlated is,
in all cases, smaller than one percent.

The correlation between $P_{\rm{mean}}$ and the EWs of the Na~I~D and
Ca~II~H\&K lines of the hosts is poorer (see top panels in Figure~\ref{fig:fig3}). 
It is more of a triangular region where at low polarization values, little or no Na~I~D lines are found, 
but at higher polarization values a wide range Na EW is found. 
For the Na~I~D lines, linear regressions provide correlation coefficients of $0.24$ for the complete
sample and $0.56$ when SN~2006X is excluded, while for the Ca~II H\&K lines the results
are $0.37$ and $0.57$, respectively.
The correlation coefficients when SN~2006X is included are not very significant.
The probability of obtaining them by chance if the EWs of lines and polarization were not correlated
are 41\% and 22\% for the Na~I~D and Ca~II H\&K lines, respectively.
When SN~2006X is excluded they become marginally significant, with probability of being obtained by chance
of 6\% and 7\% for the Na~I~D and Ca~II H\&K lines, respectively.
The overall shape of the figures suggests that there are upper and lower envelopes that limit 
the observed polarization given an EW of foreground lines of either Na~I~D or Ca~II H\&K.
The correlation between the EWs of the Na~I~D and Ca~II~H\&K lines for each SN is good, as expected
for normal interstellar media
(bottom panel in Figure~\ref{fig:fig3}).

The relation between $R_{V}$ and the mean polarization appears to be more complex.
This is presented in Figure~\ref{fig:fig4} as the $R_{V}$ estimates given in Table~\ref{tab_phot} plotted
against $P_{\rm{mean}}$.
SNe with low values of $P_{\rm{mean}}$ and small EWs of Na~I~D, display a wide range of $R_{V}$ values, although
they tend to be smaller than the typical Galactic value.
SNe with high polarization ($ \gtrsim 1\%$) and large EWs of Na~I~D appear to scatter around  a low constant value ($R_{V} \sim 1.5$). 
Taking the weighted mean of the ten $R_{V}$ in Table~\ref{tab_phot}
that have a reported uncertainty, one obtains $R_{V} = 1.50 \pm 0.06$ for the sodium sample (shown by a red horizontal line in Fig.~\ref{fig:fig4}). 
In Fig~\ref{fig:fig4} we have included the $R_{V}$ estimates given by the relation
$R_{V} \simeq 5.5 \lambda_{\rm{max}}$
\citep[][]{Serkowski1975ApJ...196..261S}.
They are plotted with black stars and qualitatively follow the same trend as the photometric estimates: a scatter at
low polarization and a convergence to $R_{V} \sim 1.5$ at larger polarization.

\section{Discussion}\label{discuss}

We chose to study the properties of the sodium sample presented in Fig.~\ref{fig:fig1}, 
because we noted that the steepness of the slope of polarization with wavelength increased with
the EW of Na~I~D lines, and the highly polarized cases could not be described by a standard Milky Way ISP. 
The preponderance of the Na~I~D lines at the redshifts of the hosts and the orientation of the polarization
pseudo vectors indicate that most of the reddening and polarization in these SNe originates in their own galaxies.
Also, fitting a Serkowski ISP law to eight of these twelve SNe requires an average wavelength of maximum polarization
$\lesssim$350~nm, a low and fairly unusual value by Milky Way standards.
On the other hand, the observed continuum polarization of the SNe in the non-sodium 
sample could be reasonably well fitted with a Serkowski-type ISP with wavelength
of maximum polarization close to 550~nm, the typical value for the Milky Way.
This, together with the lack of Na~I~D lines at the redshifts of the hosts, suggests that most of the reddening and polarization
in this group is produced in the Galaxy.
The host foreground Na~I~D lines and the unusual ISP make these twelve SNe in the sodium sample special with respect to the other seven.

The emphasis in this work is on the relationship of the continuum polarization to dust properties and it is fairly clear that for the most polarized objects,
those on the top panels of Figure~\ref{fig:fig1}, dust polarization is the dominant one.
The objects with smaller continuum polarization, however, will be more sensitive to putative intrinsic signals.
This is compounded by the fact that the latest epoch of the seven SNe with smallest continuum polarization in the sodium sample were obtained between eight and
eighteen days after maximum light, that is, at times early enough to suspect that intrinsic components could be present (see Figure~\ref{fig:EvolPmeanNaD}).
The case of SN~2011ae was already noted in Section~\ref{results} and treated especially.
Another worth mentioning is SN~2007af.
This event is an outlier both in the wavelength of maximum polarization and the orientation of the polarization angle.
Also, the continuum polarization rises towards the red up to $\sim$700~nm.
An increase in continuum polarization starting at $\sim$6500~nm and growing towards longer wavelengths was observed in spectropolarimetric models by \citet{Wang1997ApJ...476L..27W},
and was explained as the result of decreasing line blending toward longer wavelengths.
This behavior was observed by \citet{Howell2001ApJ...556..302H} in SN~1999by.
The continuum polarization of SN~2007af, however, is different from that of SN~1999by in that it does not show the break in slope expected on theoretical grounds
but rises monotously from the blue side of the spectrum.
The peculiarity of SN~2007af may be due to an intrinsic signal or to aligned foreground dust with a size distribution biased towards larger grains.
In general, contributions from intrinsic continuum polarization at a fraction of a percent level in some of our SNe cannot be ruled out.
Nevertheless, from the standpoint of this work, they imply an increase in the scatter that is more noticeable at the low polarization levels in Figures~\ref{fig:fig2} to \ref{fig:fig4}.
A more involved analysis of individual SNe will be done in forthcoming papers.

\subsection{Polarization by foreground dust}

As shown in Section~\ref{results}, the continuum polarization of the SNe in the sodium sample can be
reasonably well matched by a Serkowski law with his original constant $K$ and
different values of $\lambda_{\rm{max}}$ which do not generally follow the Whittet relation
(eq. \ref{eq:Serkowski-Whittet}).
Eight of them require $\lambda_{\rm{max}} \lesssim$0.36~$\mu$m, where two extreme cases, SN~2014J and 2002bo, were
fitted with a $\lambda_{\rm{max}} \sim$0.22~$\mu$m.
On the other hand SNe~2011ae, 2002fk and 2007af, require 
$\lambda_{\rm{max}}\sim$ 0.45, 0.55 and 0.75~$\mu$m, respectively.
Failures of the Whittet relation have been reported earlier by
\citet{Hough1987MNRAS.227P...1H} in their study of the broad--band polarization of
SN~1986G, which appeared behind the dust lane of Centaurus A and by
\citet{Patat et2009A&A...508..229P}, while \citet{Patat2015A&A...577A..53P} studied in detail
the mismatch for the cases of SNe 2006X, 2008fp and 2014J.
Non--compliance with the Whittet relation could be a general property of regions with short $\lambda_{\rm{max}}$.

An ISP model matched by a Serkowski law strongly suggests that the observed continuum polarization
originates in selective extinction by anisotropic dust grains oriented by a large scale
magnetic field in the circumstellar/interstellar medium of the parent galaxies.
This conclusion is consistent with the observed polarization being constant in time,
and with angles of polarization that tend to align with major large scale features
of the hosts galaxies projected at the SN position (see Fig.~\ref{fig:Host_Na}).

The close correlation between the observed slope $b$ and $P_{\rm{mean}}$ can be understood as the result of
polarization by regions of aligned dust, organized by the same large scale magnetic field, if $\lambda_{\rm{max}}$
is located to the blue of the $[420,580]$nm wavelength range.
In these conditions, $P_{\rm{mean}}$ will closely follow the behavior of Serkowski's $P_{\rm{max}}$ increasing
together with the slope steepness as the optical depth of aligned anisotropic dust grows.
Particular configurations or arrays of dust clouds can affect both $P_{\rm{mean}}$ and $b$ in different ways,
as mentioned earlier for the possible relations between $P_{\rm{max}}$ and $b$.
Let us consider, as illustration, two dusty regions located at different distances from the SN.
Assume that region~1 imprints a foreground polarization $P_{\rm{1}}$ due to elongated dust grains of characteristic size
$\lambda_{1}$, aligned with the local magnetic field.
If the more distant region~2 is composed of dust with the same characteristic size, aligned by the same magnetic field,
the total polarization $P_{\rm{mean}}$ will be increased ($P_{\rm{mean}} > P_{\rm{1}}$),
but the peak of polarization will remain at the same wavelength $\lambda_{1}$. 
But if region~2 contains dust grains aligned by the same magnetic field, but with a different characteristic size $\lambda_{2} > \lambda_{1}$, the total polarization $P_{\rm{mean}}$ still increases but the maximum polarization efficiency
will be shifted to wavelengths longer than $\lambda_{1}$.
If the light passes through several regions with different types of dust,
and different orientations of the organizing magnetic fields, the net effects will be depolarization and rotation
of the polarization angle with wavelength.

The correlation between $P_{\rm{mean}}$ and $b$, and the lack of rotation of the polarization angles,
suggests that most of the polarization is imprinted in few dusty regions threaded by the same large scale magnetic field.
The polarizing dust in these regions must have similar properties, resulting in similar, typically small, values of $\lambda_{\rm{max}}$.
Under these conditions, the polarization would be approximately described by a single Serkowski law \citep{Patat2010A&A...510A.108P}
and the correlation arises naturally.

\subsection{Polarization and Photometry}

As expected for polarization produced by aligned dust on Type~Ia SNe, whose intrinsic colors near maximum light fall
within a narrow range, there is a connection between $P_{\rm{mean}}$ and color, color excess and $A_V$
(Figs.~\ref{fig:Pmean_BVmax}, \ref{fig:Av_Pmean} and \ref{fig:E_BVvsPmean}).
These connections are broadly similar to the mean relations between polarization and other
photometric signals of dust found in the Galaxy. 
The tight correlation between SN color at maximum and $P_{\rm{mean}}$ in
Fig.~\ref{fig:Pmean_BVmax}, especially when SN~2006X is removed from the fit,
is expected.
Normal Type~Ia SNe have a fairly uniform blue color near maximum light so most of the
observed spread in color ought to be due to extinction.
Note that SN~2006X appears as the discrepant case in all of these fits, and that it
always displays a polarization too high in comparison with the value that the
correlation of the rest of the objects would suggest.
The highly reddened SNe 2006X and 2014J appear to be indicative of the range of expected polarization values given a reddening.

Regarding the color excesses, Fig.~\ref{fig:E_BVvsPmean} repeats the message in
Fig.~\ref{fig:Pmean_BVmax}.
The sodium sample shows a correlation between color excess and polarization when
SN~2006X is excluded.
\citet{Serkowski1975ApJ...196..261S}, \citet{Whittet1978A&A....66...57W} and \cite{Clayton1988ApJ...327..911C} among other authors have studied
correlations between color excess and $P_{\rm{max}}$ of Milky Way stars.
With the exception of SN~2006X \citep[see][]{Patat2015A&A...577A..53P}, we cannot provide a reliable measurement of
$P_{\rm{max}}$ for our SNe since it typically falls at the blue edge of our spectra, or beyond.
If we assume an average $\lambda_{\rm{max}}=0.35~\mu$m for most of the sample, the good match between Equations
\ref{eq:Serkowski} and \ref{eq:straightfit} implies that $P_{\rm{mean}} \simeq 0.86 P_{\rm{max}}$.
With this translation, we see that our observations follow the average relation found by 
\citet{Fosalba2002ApJ...564..762F} in their statistical study of polarized and reddened
stars in the Milky Way (see in particular their Eq.~3).
Also, a noted result from \citet{Serkowski1975ApJ...196..261S} is that for a given
color excess there is a limit to the amount of polarization that the dust can imprint.
This is given by the inequality $P_{\rm{max}} \leq 9.0 E(B-V)$. This limit,
translated to $P_{\rm{mean}}$, has been plotted
in Fig.~\ref{fig:E_BVvsPmean} with a magenta solid line.
The figure shows that
all the SNe with $E(B-V) \gtrsim 0.1$, including our discrepant highly--polarized SN~2006X,
are well below the Galactic upper limit.
According to the standards of polarized stars in the Milky Way, Type~Ia SNe display modest
polarization levels given their color excesses.

The correlation between $P_{\rm{mean}}$ and $A_V$ shown in Fig.~\ref{fig:Av_Pmean}
replicates in part the correlation that the former has with $E(B-V)$.
In this case, the correlation is good both with and without SN~2006X but improves when
it is removed from the fit.
Again, SN~2006X appears as too highly polarized for the given extinction.
If a value of $R_V$ is assumed, Serkowski's upper limit of polarization for a given color excess
can be transformed to an upper limit for a given extinction.
As a guide to the eye we have plotted the upper limit given by $P_{\rm{mean}} \leq 2.5 A_V$, assuming $R_V \sim 3.1$.
As shown in Figure~\ref{fig:Av_Pmean}, some SNe exceed this limit, indicating that the assumed value of $R_V$
is larger than the actual value appropriate for the SNe.
An $R_V \sim 2.0$ is a better upper limit to Fig.~\ref{fig:Av_Pmean} and it also agrees better with photometric
studies of large SN Ia samples that tend to find values of $R_V$ smaller than those typical of the Milky Way
\citep{Wang2005ApJ...635L..33W,Goobar2008ApJ...686L.103G,Folatelli2010AJ....139..120F,Burns2014}.

In general, the comparison of Figs.~\ref{fig:Pmean_BVmax}, \ref{fig:Av_Pmean} and
\ref{fig:E_BVvsPmean} with
Figure~9 of \citet{Serkowski1975ApJ...196..261S} suggests that our SNe display a noisy correlation.
The dispersion in polarization increases with polarization so that the points scatter in a  wedge.
There appear to be upper and lower boundaries between polarization, color, color excess or extinction,
rather than an upper limit for the polarization for a given color, color excess, or extinction,
especially for those objects with $E(B-V) \gtrsim 0.2$.
This indicates that the efficiency of dust alignment is similar
in the foreground regions of these SNe, particularly at moderate and low-extinction sight lines.

Spectropolarimetry can also enrich the discussion on the value of $R_V$ through its connection
with $\lambda_{\rm{max}}$.
Many previous studies have suggested that the value of $R_{V}$ measured for Type Ia SNe
is generally smaller than the typical value $\sim3$ of our Galaxy and our
spectropolarimetry is qualitatively consistent with them.
We have plotted in Fig.~\ref{fig:fig4} the values of $R_V$ expected from the
relation $R_{V}/\lambda_{\rm{max}} \sim 5.5$ found by \citet{Serkowski1975ApJ...196..261S}
for the Milky Way (see the black stars).
For fixed $K$, $\lambda_{\rm{max}}$ essentially determines the slope of the Serkowski law
(Eq.~\ref{eq:Serkowski}) in the wavelength range used to fit Equation \ref{eq:straightfit},
and hence the slope $b$ of our fitted straight line.
In Fig.~\ref{fig:fig1} we distinguish three groups of foreground polarization slopes:
$b \lesssim 0, b\sim0$ and $b\gtrsim0$.
They are a consequence of the three ranges of $\lambda_{\rm{max}}$ mentioned earlier for
the twelve SNe in the sodium sample
($\lambda_{\rm{max}}\sim$0.20-0.36~$\mu$m for eight SNe, $\sim0.50-0.55~\mu$m for three of them, and
$\sim0.75~\mu$m for the remaining one) which, in turn, imply $R_{V} \sim $ 1.8, 2.8 and 3.9, respectively.
Although the values of $R_{V}$ suggested by the observed polarization are qualitatively
consistent with the values estimated from photometric observations, the quantitative differences
could be an indication that the relation $R_{V} \sim 5.5 \lambda_{\rm{max}}$ does not apply
to the foreground polarizing dust of these Type~Ia SNe host galaxies.

The polarimetric measurements suggest that a relevant fraction of the foreground dust in all the SNe of the sodium sample
is located in a fairly stable environment that has had time to relax.
Turbulent or shocked regions will not permit a correlation between color excess and polarization.
The polarizing dusty regions cannot be close to star-forming regions, nor 
in a circumstellar region produced by a recent episode of mass ejection by the SN progenitors,
nor in the CSM shocked by the SN ejecta.
The previous conclusion rules out a polarization imprinted by a local CSM.
This has been confirmed in the case of SN~2014J by high-resolution spectroscopic studies by \citet{Maeda2016ApJ...816...57M},
who find that the sodium absorption systems originate on interstellar scales.

Still, even if not in the CSM realm, the polarizing dust could be associated with the SNe or their progenitors.
Studies of polarization of transmitted light in stars where at least part of the foreground
matter can be seen, as in reflection nebulae, show that when the nebulae appear relaxed,
they display well ordered physical structures, generally regularly spaced and approximately
parallel filaments, which are also parallel to the polarization pseudo-vectors that trace
the global Galactic structures.
This indicates that there are regions, which could be called distant CSM, which are already
threaded by the global magnetic field
\citep{Hall1963bad..book..293H,Spitzer1968ITPA...28.....S,Marraco1988LNP...315..120M}.
The required condition is that the dusty gas stays under the influence of the global
magnetic field for a time longer than the time scale for dust alignment, $\tau_{DA}$.
According to \citet{Draine_Weingartner_1997_ApJ_480_633} this time scale is of the order
$2\times10^5 \lesssim \tau_{DA} \lesssim 1\times10^7$~years.

\subsection{Polarization and Foreground Interstellar Absorption Lines}

\citet{Sternberg2011Sci...333..856S} studied the velocity structure of the
Na~I~D systems in 35 SNe Ia and categorized them in three classes:
blueshifted when they display a dominant feature with one or more minor
components at shorter wavelengths;
redshifted when the minor components appear at longer wavelengths from the dominant feature;
and single/symmetric when there is a single feature, or a system of lines
symmetrically centered at the expected velocity of the SN
host\footnote{We note that in the single case it is really
impossible to establish a Doppler shift for the line, since there is not a main component to adopt as rest frame. 
It is not possible to establish whether the single Na feature is shifted or not.}.
They find a statistically significant excess of blueshifted Na~I~D systems in the foreground
matter of Type~Ia SNe and interpreted this as evidence of outflows from the SN progenitors.
Seven of our twelve SNe are included in their sample:
SNe~2006X, 2008fp, 2007le, 2002bo, and 2007fb, are blueshifted,
SN~2010ev has a single absorption feature, and SN~2007af is redshifted.
In order of strength of polarization signal (c.f. Fig.~\ref{fig:fig1}) the blueshifted SNe are among the highest polarization
(with the exception of SN~2007fb), the single symmetric one is in the middle and  the redshift one near the bottom, at low polarization values.

\citet{Phillips2013ApJ...779...38P} found that the Type~Ia SNe which displayed foreground Na~I~D lines had
a variety of proportions of sodium column density and $A_V$ (i.e. different dust-to-gas ratios).
Most of them would approach the average ratio found for the Milky Way, but about a third of their sample showed anomalously large Na~I~D column densities
compared with the Milky Way ratio.
The majority of these SNe with anomalously strong Na~I~D absorption are
blueshifted, according to \citet{Sternberg2011Sci...333..856S}. 

In our Fig.~\ref{fig:fig3}, the EWs of the Na~I~D and Ca~II~H\&K foreground absorption lines
show a generally similar trend with polarization, although
the correlation is not as good as those between the polarization
and color, color excess, or extinction, as discussed previously.
Qualitatively, the two panels in Fig.~\ref{fig:fig3}
repeat what \citet{Phillips2013ApJ...779...38P} found for the correlation between
$A_V$ and the column densities of sodium and potassium.
Since our sample is relatively small, we refrain from defining a region of average or
normal values of EW for a given $P_{\rm mean}$ and measure departures from those
normal values as excess.
The plots can, nevertheless, be interpreted as a correlation where the observed
EWs of Na~D~I or Ca~II H\&K lines fall within a range with upper and lower limits, which diverge from the mean
as $P_{\rm mean}$ increases (a wedge-like scatter plot).

Our observations can be understood under the assumption that the dominant source of polarization and
extinction in the foreground of these SNe consists of a few regions of gas and dust,
permeated by the same large scale magnetic field, where a significant fraction of the dust
has smaller average size than typical in the Milky Way.
The regions could be close to the SN, in what we may call distant CSM,
but the dominant part of the foreground matter is not in what is usually
understood as CSM, under the direct mechanical or radiative influence of the SN explosion.
It is reasonable to expect that, if not perturbed, this mix of dust and gas regions will
evolve towards an equilibrium-like state, with a fairly well defined dust-to-gas ratio.
These average ratios give rise to the mean correlation between gas column density
and $A_V$ in the Milky Way and other galaxies, found by \citet{1993A&AS..100..107S}
and \citet{Phillips2013ApJ...779...38P}.
In some of the sight lines towards our Type~Ia SNe, however, the equilibrium-like stage has
either not been reached or has been perturbed, resulting in a relative excess of Na~I gas with respect to dust.
A scenario like this would explain the enhanced strength of the Na~I~D lines in some SNe given
their polarization (as in Fig.~\ref{fig:fig3}), or their $A_V$ \citep[as in][]{Phillips2013ApJ...779...38P}.

Where is the polarizing dust? It could be anywhere along the light path, although
the correlations between polarization, reddening, and EW at the redshift of the parent galaxy
mean that a major fraction is within the SN host galaxies.
In the context of \citet{Sternberg2011Sci...333..856S}, 
it is tempting to imagine the dust close enough to the SNe so as to connect their progenitors, to the presence of,
excess of, and/or blueshift in, the Na~I~D absorption lines, but it cannot be very close to the SNe because 
the grains would be evaporated by the UV radiation.

As mentioned earlier, a small percentage of Type~Ia SNe show variability in the EW of the Na~I~D lines
\citep[SN~1999cl, 2006X and 2007le,][respectively]{{Blondin2009ApJ...693..207B},Patat et2009A&A...508..229P,Simon2009ApJ...702.1157S}.
This indicates that some of the CSM responsible for the Na~I~D absorption is actually within the radiative influence of the SN explosion.
The cases of SNe 2006X and 2007le, which were studied at high resolution, reveal that the variable Na~I~D components imply
a small fraction of the total EW.
This fact, together with the constancy of continuum polarization in time and the realization that these SNe generally
follow the same trends as the others in the correlations shown in Figures~\ref{fig:fig2} to \ref{fig:fig4},
lead us to conclude that the variable Na~I~D components are associated with polarizing dust.

Matter in the line of sight towards three events in our sample has been detected through light echoes (LE).
\citet{Wang2008b...677.1060W} report the discovery of a light echo in SN~2006X detected approximately 300 days after maximum.
They locate the reflecting dust between 27 and 170~pc from the SN, and also propose the existence of a second, inner, LE.
\citet{Crotts2008ApJ...689.1186C}, using observations taken at 680 days after maximum,
confirm the more distant LE, constraining its distance to 26~pc, and rule out the possibility of the inner one.
In addition, they claim that the reflecting dust structure is the dominant source of extinction in SN~2006X. 
\citet{Drozdov2015ApJ...805...71D} report two LEs in SN~2007af discovered in data taken three years after maximum.
There is a main outer echo located at $\sim790$~pc and produced by a foreground dust sheet which should also dominate the extinction.
The second, inner LE, would be between 0.45 and 90~pc, depending on whether the dust is in front or behind the SN.
\citet{Crotts2014arXiv1409.8671C} reports the detection of a LE from SN~2014J at 300~pc and a possible inner one at 80~pc from the SN.
Finally, \citet{Yang2015arXiv151102495Y}, use HST observations taken $\sim$277 and $\sim$416 days after $B$ maximum to show
that SN~2014J develops light echoes from CSM located at foreground distances of $\sim$222 and $\sim$367 pc, with small--scale
structures reaching down to approximately one pc.
They also point out that the amount of dusty material involved in the echoes is much smaller than that involved in the total
extinction.
In all cases, the echos are bluer than the SNe, indicating high scattering efficiency at short wavelengths.
This relates to a small average dust grain size, in agreement with our observations of small $\lambda_{\rm{max}}$.
It is difficult to associate the presence of LEs in SN~2006X and 2014J with other obvious peculiarities in their 
polarization.
They are two of the most reddened and polarized SNe of the sodium sample but follow the general trends of the rest.
The case of SN~2007af is different. It is one of the SNe with lowest polarization, the angle of polarization is not aligned
with major features of the parent galaxy, and the wavelength of maximum polarization is the longest of the sample.
\citet{Drozdov2015ApJ...805...71D} found evidence for two light echoes.
One of them, the outer echo, is located some 790~pc from the SN, in the foreground.
The measured optical depth suggests that the dust in it is responsible for most of the extinction of the SN light curves, and,
hence, it should also dominate the ISP.
The dust responsible for the other, inner, echo is closer to the SN but its location could not be unambiguously determined.
It could be $\sim$90~pc in the foreground, or 0.45~pc in the background.
Even at the closest distance, the delay between the scattered light and direct light is far longer than the range of epochs at which we
observed the SN.
We conclude that the matter causing the light echoes in SN~2007af is not responsible for polarizing the light at early times.

Could matter from the SN progenitors be at distances comparable to those inferred from the LEs {\em and} polarize light?
The time scale for alignment of the dust in the magnetic field allows us to put some constraints, albeit weak, on the location.
The dust has to be in a fairly relaxed region, far from strong shocks, strong winds, or other
large scale hydrodynamic phenomena, and it must have remained in this unperturbed state for a time longer than $\tau_{DA}$.
In the SD scenario, the dust could have originated in the winds of an evolved companion.
At the typical velocity of 100~km~s$^{-1}$, the smallest $\tau_{DA}$ given by
\citet{Draine_Weingartner_1997_ApJ_480_633}, $2\times10^5$ years, will put the matter
at some 20~pc from the progenitor.
Another limit comes from the constancy in time of the foreground polarization.
Taking the longest time period covered by spectropolarimetric observations of a SN in our sample, 7 weeks in the case of SN~2006X, as representative of
the sample and 2$\times 10^4$~km~s$^{-1}$ as a typical maximum expansion velocity of
SN ejecta, means that the dust has to be farther than $\sim$0.1~pc in order to be safe
from the ejecta colliding with it.
Winds from the progenitor expanding during times as long as $\tau_{DA}$, or longer,
could put matter with the potential to provide aligned dust at a safe distance from the
SN explosion, but it will not be close enough to be called CSM.
The numbers are within the range of possibilities for the light echoes observed but are only
marginally consistent with the scenario proposed by \citet{Soker2014}, where the foreground matter causing
the enhanced Na~D~I lines has to be at distances of $\sim 10^{18}$~cm (0.32 pc) for the light of the SN to cause desorption of sodium atoms
from the dust grains.

Why are the galactic environments in the foreground of the SNe with host Na~I~D lines in absorption
systematically different from what we see in similarly reddened stars in the Galaxy?
Why do they show more efficiency in polarizing blue light than those towards SNe that do not display these narrow lines?
It may be relevant to note that there are lines of sight in our galaxy that show a relative increase of polarization in
ultraviolet wavelengths.
\citet{Martin1999} study 28 lines of sight using HST data and report that two of them have an unusually short $\lambda_{\rm{max}}$.
They claim that the maximum polarization is high but the polarization efficiency is low, and
remark that these two sight lines lie in regions of active star formation, where there is evidence of
stellar winds and supernova activity.
They claim that the statistical relevance of these two cases is small, and attribute the
difference to variations in the dust grain shapes and/or efficiency in grain alignment.
In our region of the Galaxy, short $\lambda_{\rm{max}}$ is the exception, not the norm \citep{Patat2015A&A...577A..53P}.

It is tempting to surmise that the same process that produces an excess of sodium gas produces the polarization spectrum
with short $\lambda_{\rm{max}}$.
A scenario proposed by \citet{Hoang2015arXiv151001822H} carries the promise to account for the contrasting elements of order
and chaos revealed by Figs.~\ref{fi:lambdamax-histogram}--\ref{fig:fig4}, and the results of \citet{Sternberg2011Sci...333..856S} and
\citet{Phillips2013ApJ...779...38P}.
Hoang\'{}s study of the dust grain size distribution and alignment in the foreground of SNe 2006X, 2008fp, 2014J, and 1986G, finds that reproducing the small values of $R_{V}$ requires both an increased total mass of small grains and efficiency in their alignment.
He suggests that these SNe had matter located at distances much greater than the sublimation radius of the grains, but close enough
that the radiation pressure from the SN light accelerates the dust clouds that are closer and makes them collide with those that are farther.
The collisions alter the grain size distribution and bias them towards small sizes, while the radiative torques
induced by the SN radiation promote a faster alignment, especially of the small grains.
He finds that the relevant distance scales for these processes to work are 1--20 parsecs while the time scales for alignment can be as short as
few days.
A scenario like this will also explain an increase in the gaseous phase of those atoms like sodium, which are 
generally more difficult to incorporate in the solid phase of ISM and more sensitive to photodesorption
\citep{Draine2004coun.book..213D}, and will also result in a preponderance of blueshifted Na~I~D line systems.
The SNe that \citet{Hoang2015arXiv151001822H} focused on are the four most polarized of our sodium sample.
The rest of the SNe in the sodium sample could differ just in the amount of total ISM mass within the first $\sim$20~pc or whether there are
different clouds that can be differentially accelerated.
A critique of this scenario will argue that in cases like SN~2014J high resolution spectra reveal the presence of many individual
Na~I components \citep[ten, according to][]{Maeda2016ApJ...816...57M}, which would be associated with different dusty regions at
different distances from the SN, with many of them probably very distant.
It is, then, neccesary to explain why those that are closer to the SN are the ones which dominate the polarization.
Another issue to raise in the scenario of \citet{Hoang2015arXiv151001822H} is that ISPs with
Serkowski-like polarization and small $\lambda_{\rm{max}}$ are not typically observed in the lines of sight towards extragalactic
core collapse SNe.
The general case for them is an ISP that is well fitted by a Serkowski law with standard Galactic parameters \citep[][]{2001PASP..113..920L,2002AJ....124.2506L,2007A&A...475L...1M,2011ApJ...739...41C}.
It will be also necessary, then, to explain why the radiation pressure from a SN causes such different effects in the environments of Type~Ia
and core collapse SNe.

An estimate of the physical conditions in the regions where the foreground Na~I~D, K~I, and Ca~II lines are formed will
certainly help to better constrain their location and provide firmer grounds for understanding.
Environmental tests, like the Routly-Spitzer effect \citep{Routly1952ApJ...115..227R}, will be an appropriate tool.
We cannot apply it to our data set because it requires column densities of the absorbing species, a piece of data that we cannot reliable estimate from
our low resolution spectra.

\section{Conclusions}\label{conclusions}

We have presented observations of the continuum polarization in the wavelength range from 420 nm to 580 nm of the 19 Type Ia SNe in our database that were observed at least twice, with one observation close to the time
of maximum light.
We have found that the presence of narrow Na~I~D lines of foreground origin at the redshift of the parent galaxies
is connected with special characteristics of the polarization in this blue wavelength range.
This led us to split the SNe into a sodium sample, which shows the lines, and a non--sodium sample, which does not.

\begin{figure}[t!]
\centering
\includegraphics[width=17cm]{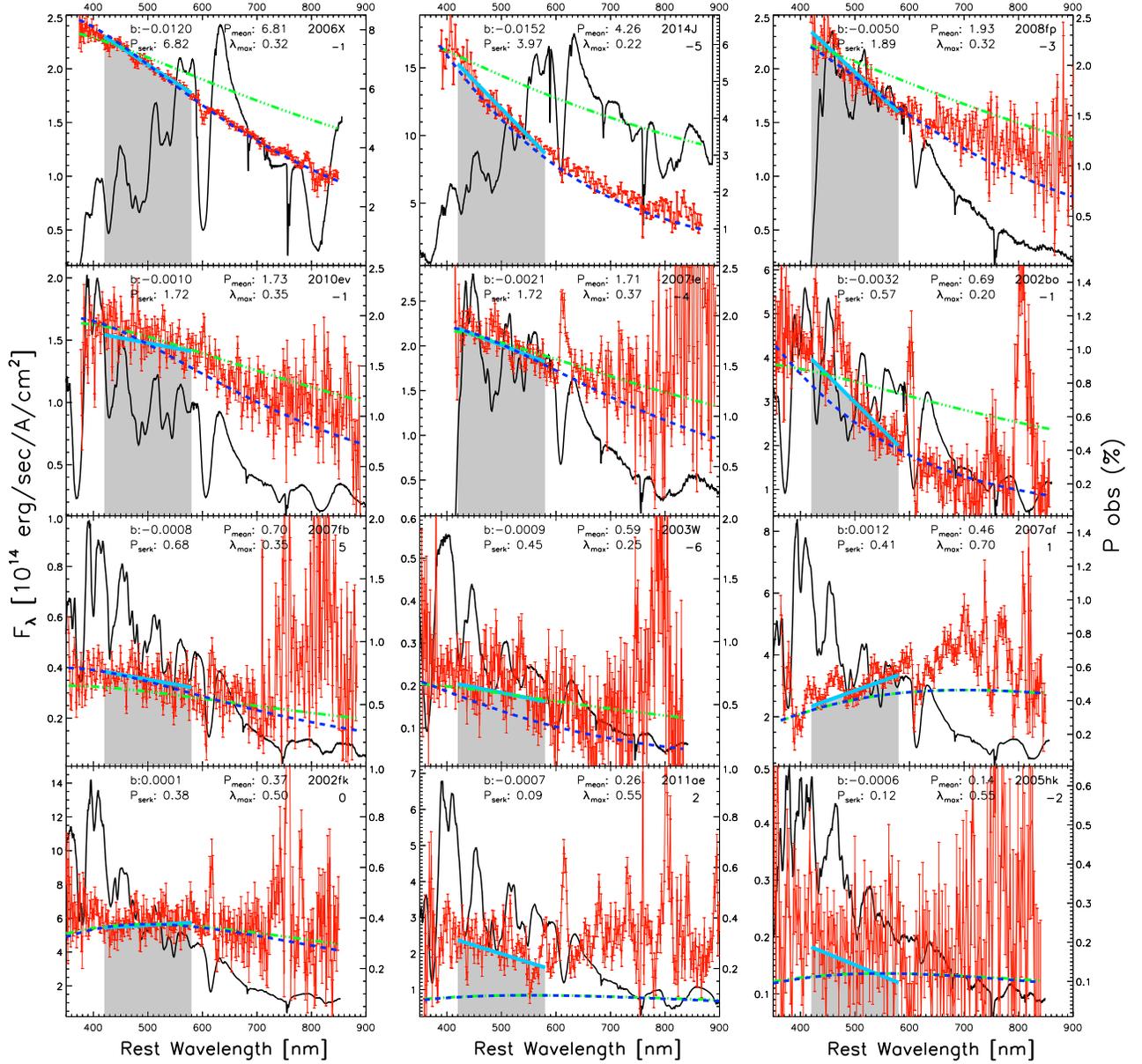}
\caption{The polarization and intensity spectra of the SNe in the sodium sample (see Table~\ref{tab_obslog}).
The left vertical axes give monochromatic flux and the right ones percent polarization. Note that the vertical
scales are different in different panels.
The polarization spectrum is given by the red solid line and the intensity spectrum by the black solid line.
The dashed blue line is the original Serkowski law 
\citep[our eq.~\ref{eq:Serkowski},][]{Serkowski1975ApJ...196..261S}.
The dot--dashed green line is Serkowski's law with parameter $K$ given by
eq.~\ref{eq:Serkowski-Whittet} \citep{Whittet1992ApJ...386..562W}.
The gray-shaded area marks the wavelength range 420-580~nm, used to fit a straight line
to the observed polarization spectrum.
The fit is plotted with the cyan line.
The parameters of the fit, slope $b$ and mean percent polarization $P_{\rm{mean}}$,
are given at the top of each panel.
From top-left to bottom-right, panels show a decreasing $P_{\rm{mean}}$ or, equivalently, decreasing
shaded area.
}
\label{fig:fig1}
\end{figure}

\begin{figure}[t!]
\centering
\includegraphics[width=17cm]{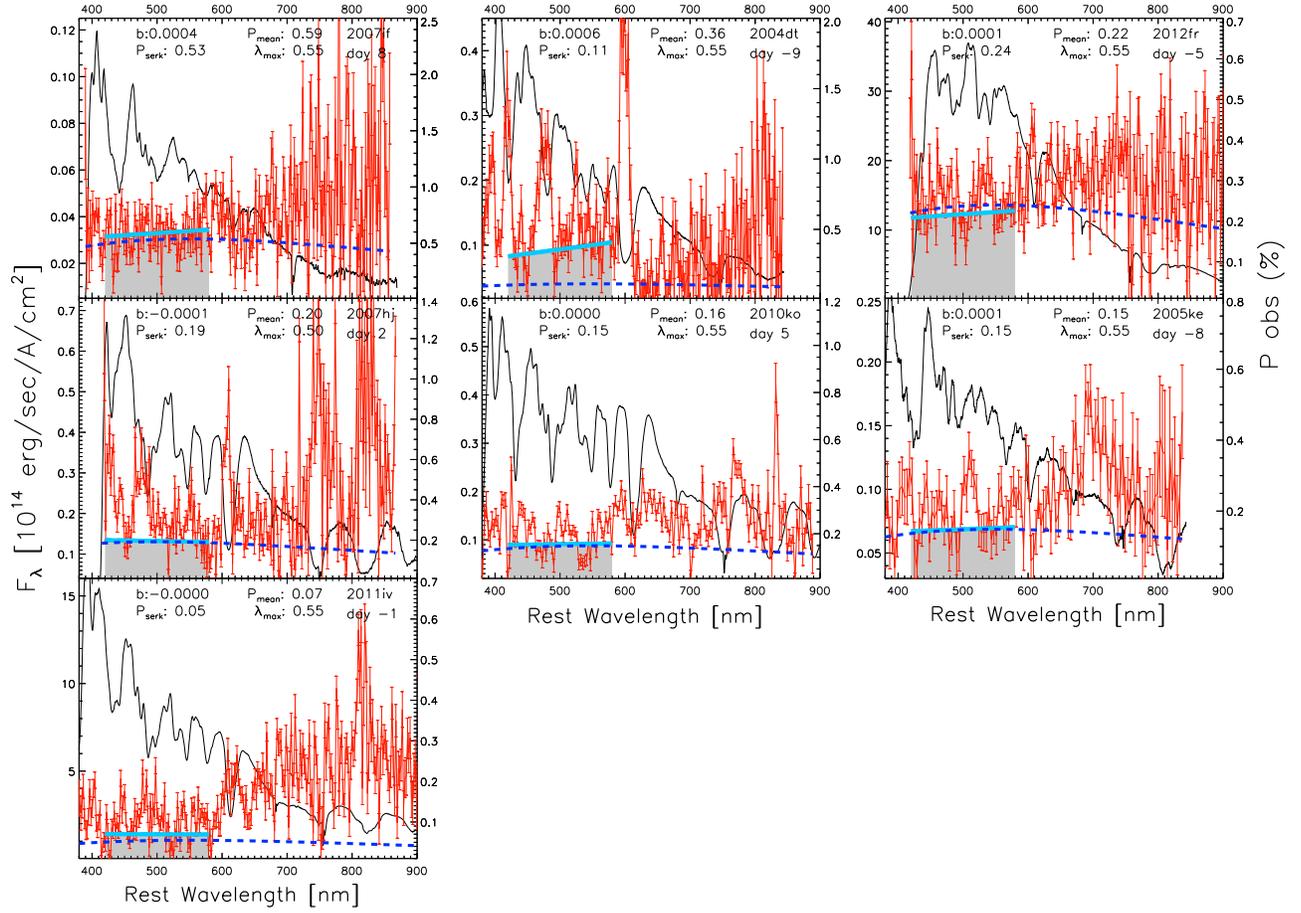}
\caption{Same as Figure~\ref{fig:fig1} for the polarization and intensity spectra of the seven
SNe that did not show narrow Na~I~D lines at the velocity of their hosts in our low-resolution
spectra.
}
\label{fig:fig1b}
\end{figure}

\begin{figure}[t!]
\center
\includegraphics[width=12cm,angle=0]{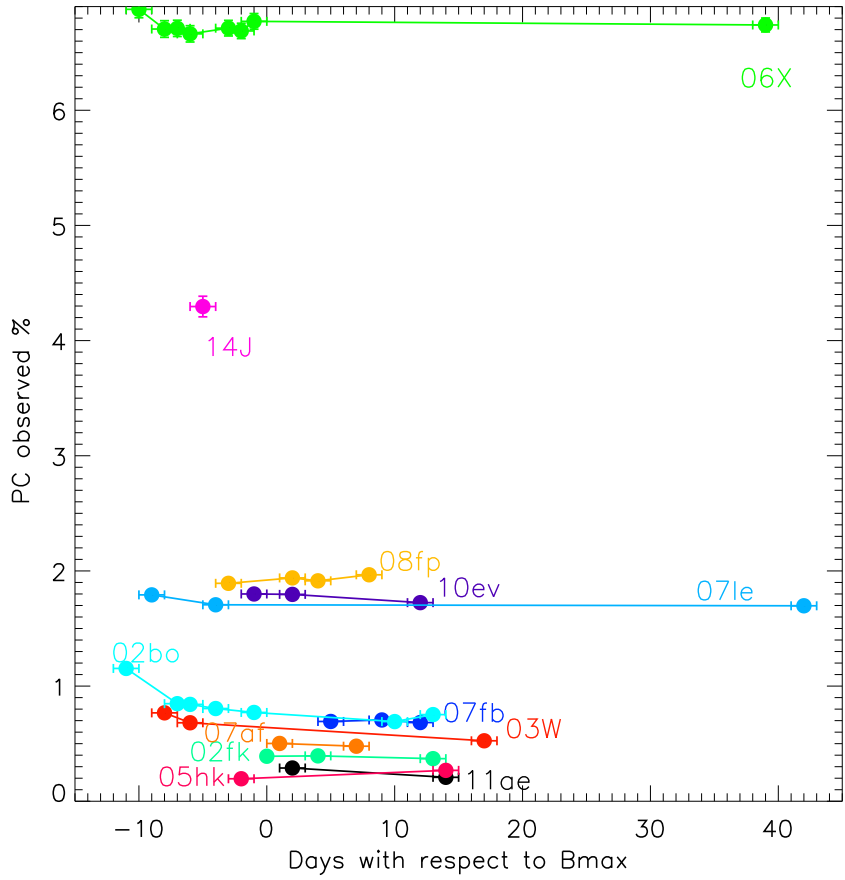}
\caption{
Evolution in time of the observed continuum polarization for the sodium sample (see text for definition of continuum polarization).
}
\label{fig:EvolPmeanNaD}
\end{figure}

\begin{figure}[t!]
\centering
\includegraphics[width=17cm]{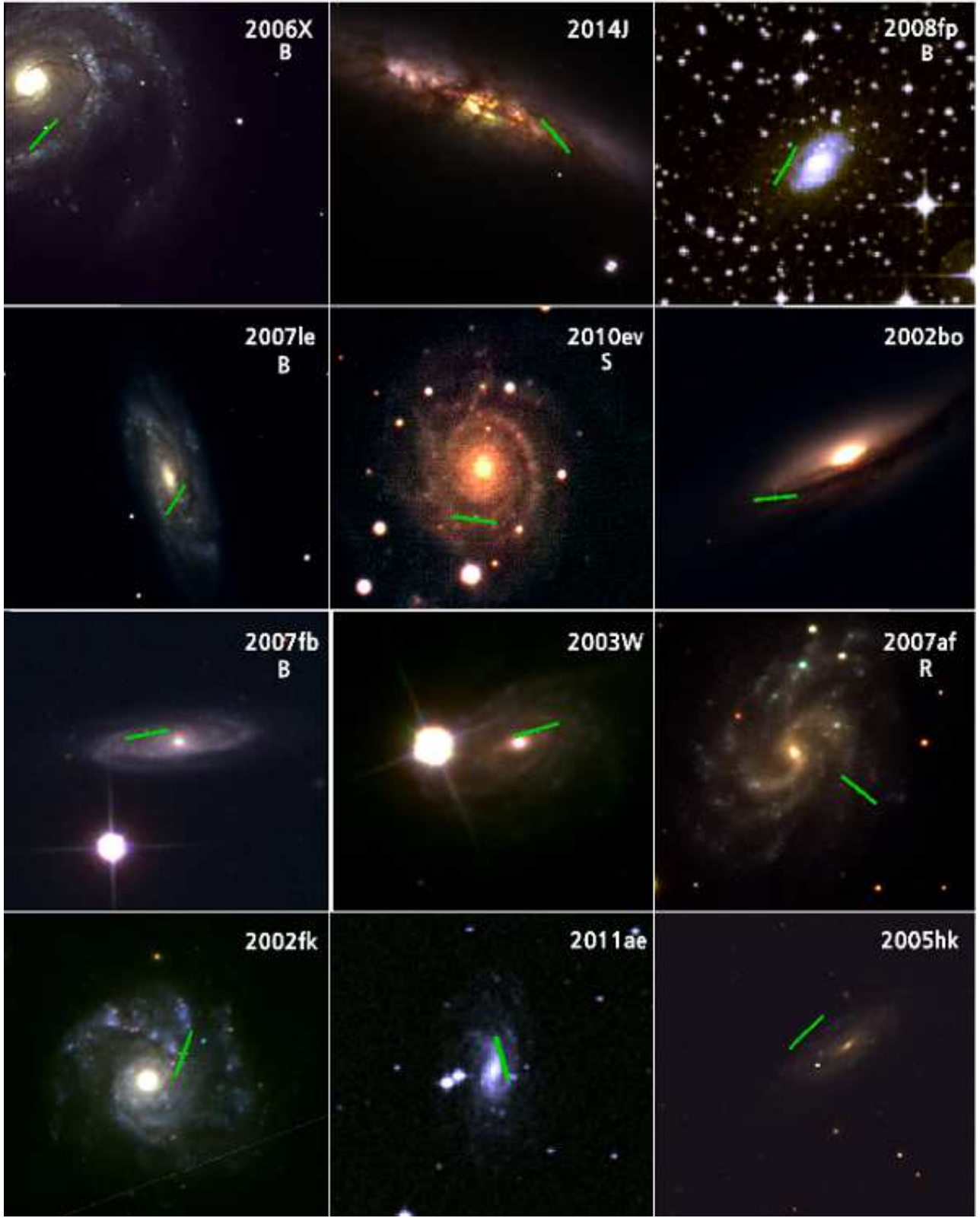}
\caption{Orientation of the polarization angle for the 12 SNe in the sodium sample. The bars are centered at the position of the SNe. North is up and East is to the left. Labels B,S and R correspond to Sternberg\textasciiacute s classification of the sodium line shift.}
\label{fig:Host_Na}
\end{figure}

\begin{figure}[t!]
\centering
\includegraphics[width=17cm]{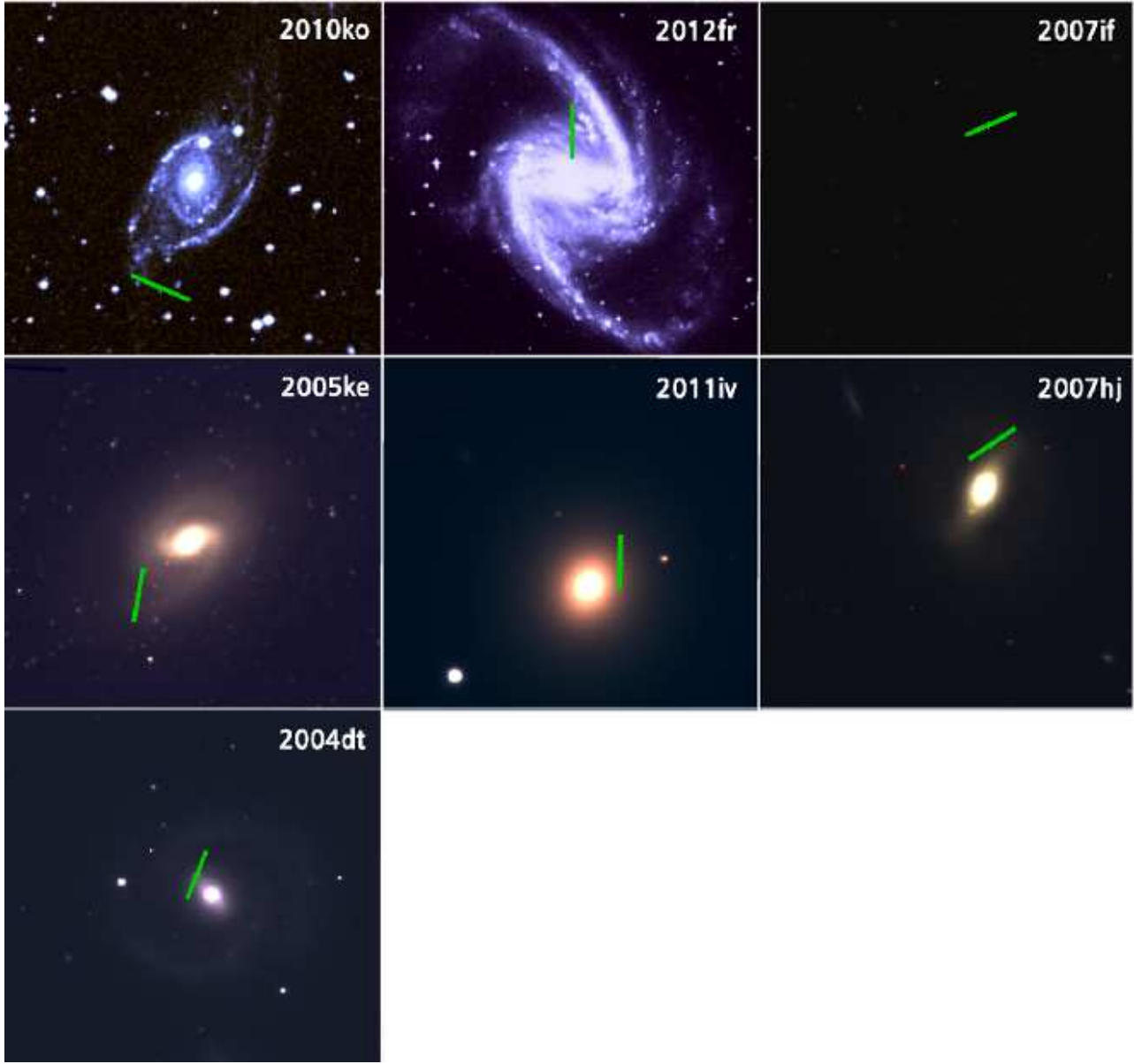}
\caption{Same as Figure~\ref{fig:Host_Na} for the seven SNe in the non-sodium sample.
}
\label{fig:Host_NoNa}
\end{figure}

\begin{figure}[t!]
\centering
\includegraphics[width=17cm]{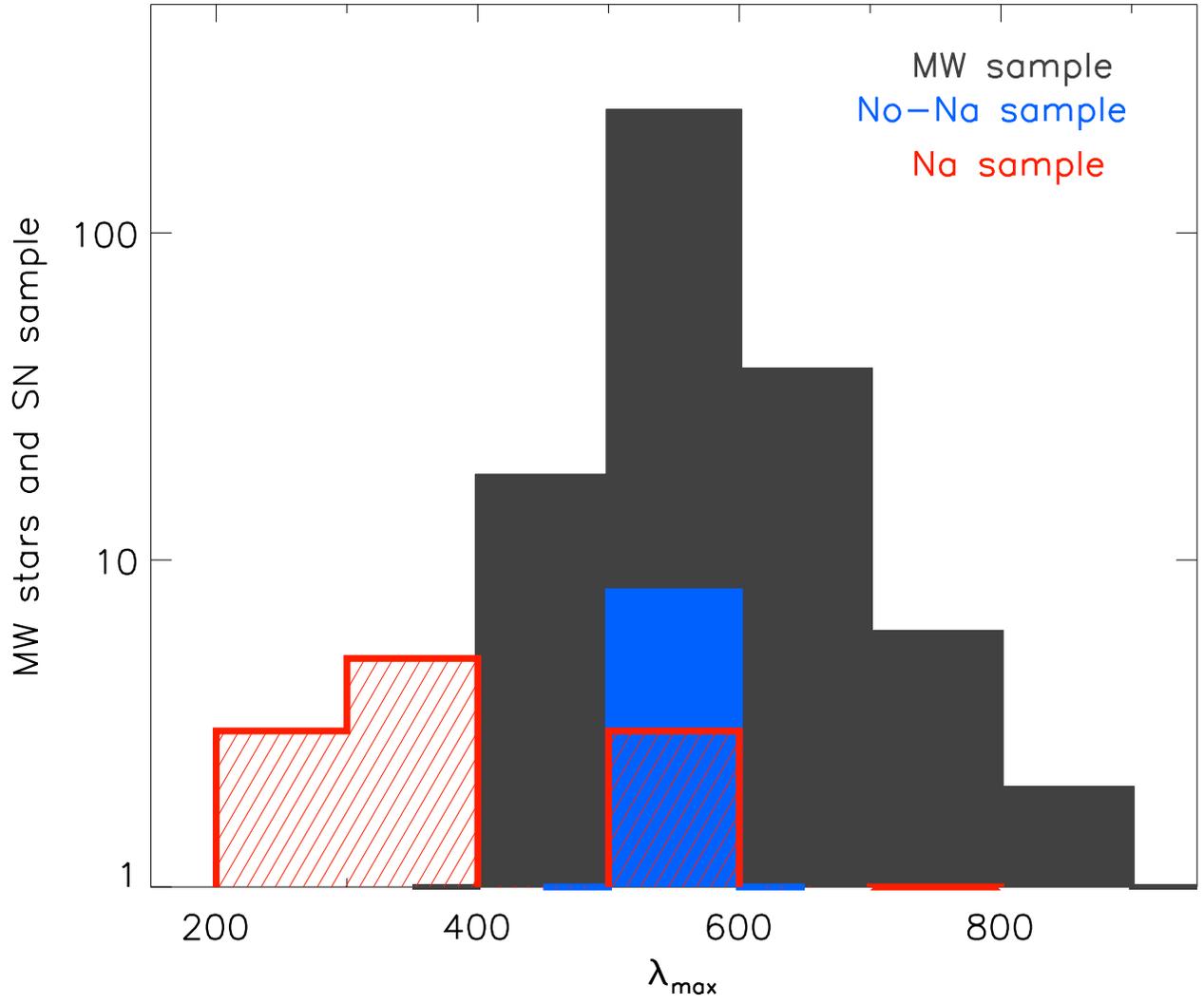}
\caption{Histogram of $\lambda_{max}$ for SNe in the sodium sample (red shaded area), the non-sodium sample (blue filled area) and stars
in the Galaxy (shaded in dark grey).
}
\label{fi:lambdamax-histogram}
\end{figure}

\begin{figure}[t!]
\center
\includegraphics[width=12cm,angle=0]{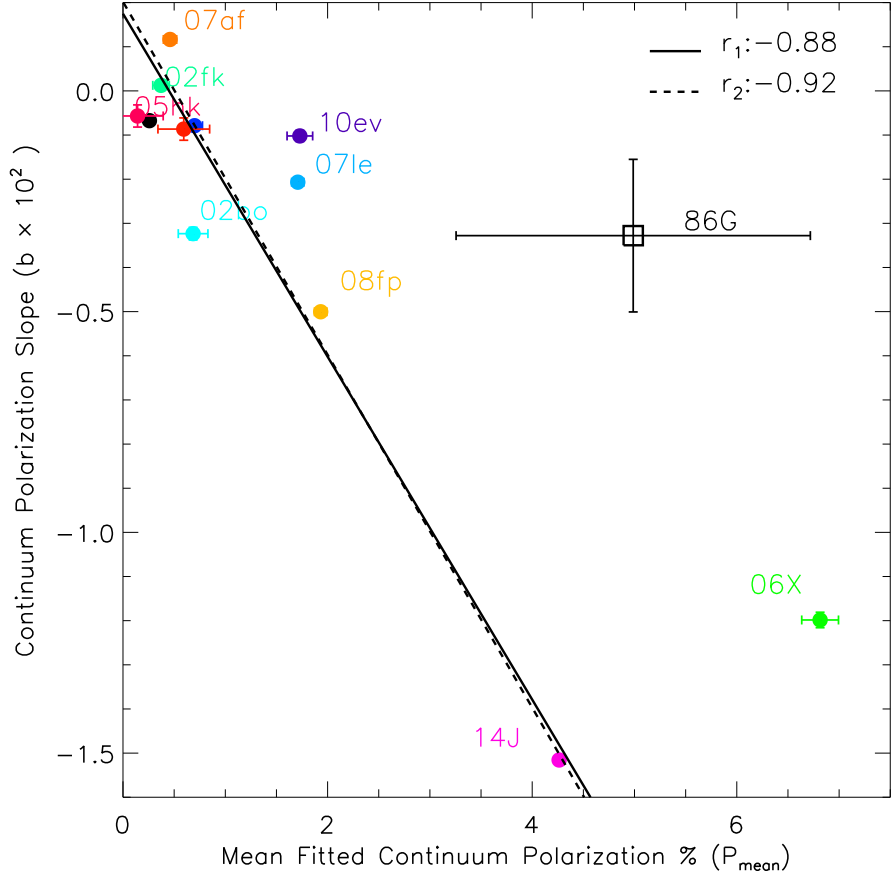}
\caption{Correlation between the parameters $b$ and $P_{\rm{mean}}$
that describe the observed continuum polarization (see text and Fig.~\ref{fig:fig1}).
The solid line shows the linear fit to all points (correlation coefficient $r_1$)
and the dashed line the linear fit when SN~2006X is excluded (correlation coefficient $r_2$).
}
\label{fig:fig2}
\end{figure}

\begin{figure}[t!]
\center
\includegraphics[width=12cm,angle=0]{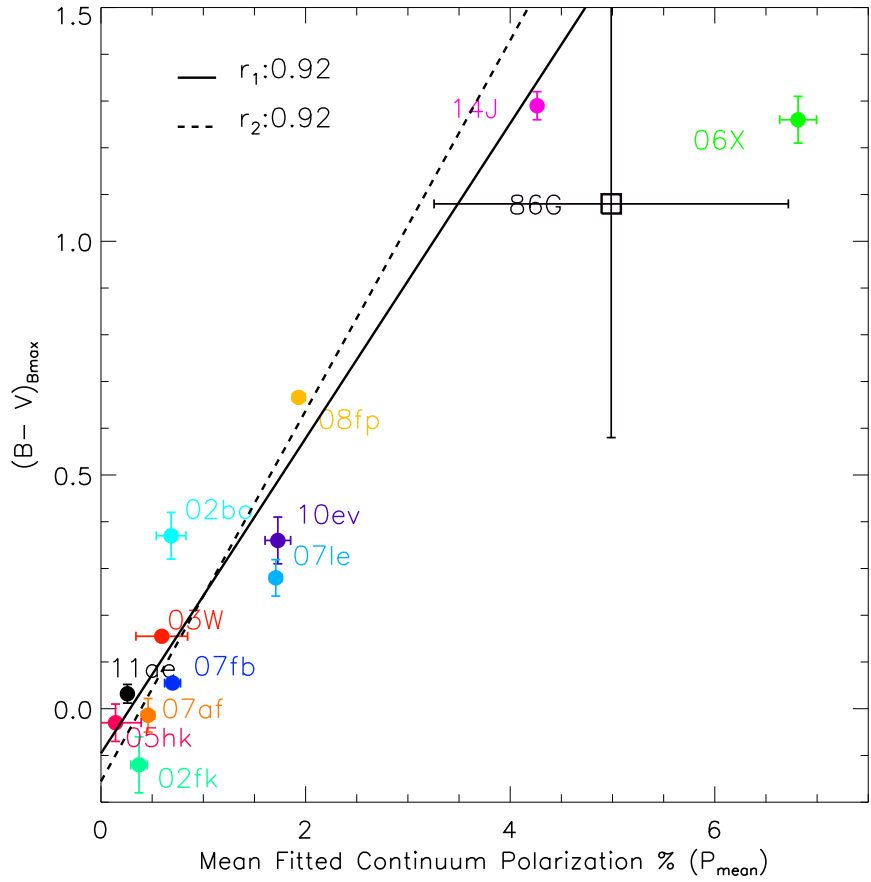}
\caption{Correlation between the observed color at maximum light and the continuum
polarization.
As in the previous figure, the solid line shows the fit to the entire sample
and the dashed line excludes SN~2006X. The correlation coefficients are r$_{1}$ and r$_{2}$,
respectively.
}
\label{fig:Pmean_BVmax}
\end{figure}

\begin{figure}[t!]
\center
\includegraphics[width=12cm,angle=0]{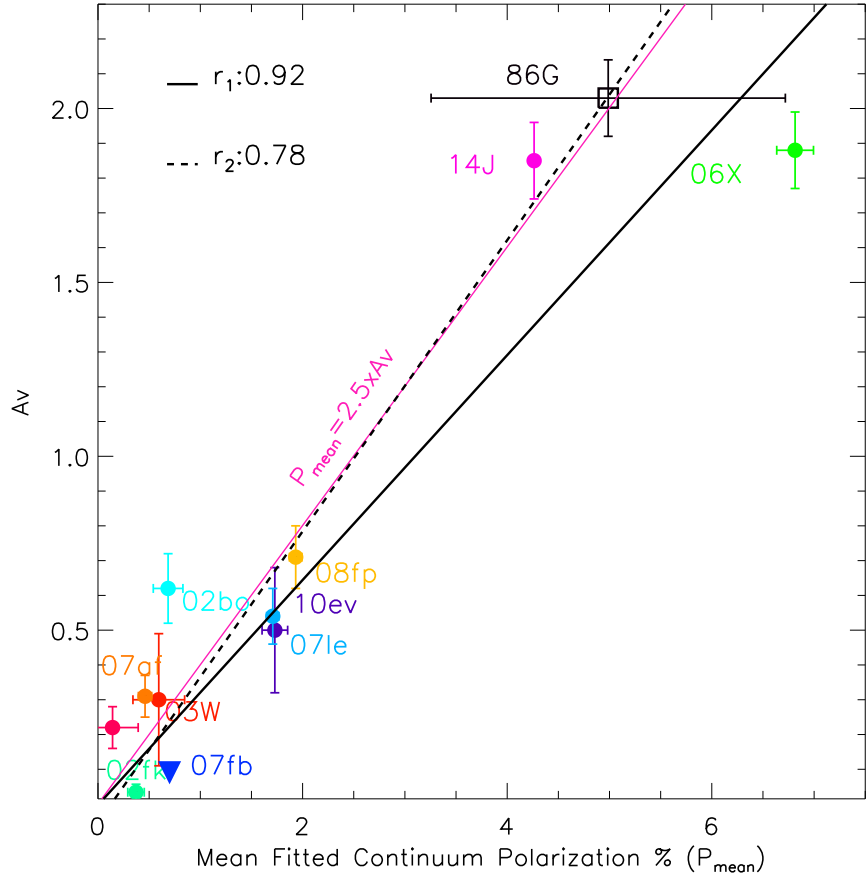}
\caption{Correlation between the extinction in the visual band and $P_{\rm{mean}}$.
The solid line shows the linear fit to all points and
the dashed line the linear fit when SN~2006X is excluded (correlation coefficients r$_{1}$ and r$_{2}$ respectively).
Down pointing triangles are upper limits to $A_V$.
}
\label{fig:Av_Pmean}
\end{figure}

\begin{figure}[t!]
\center
\includegraphics[width=12cm,angle=0]{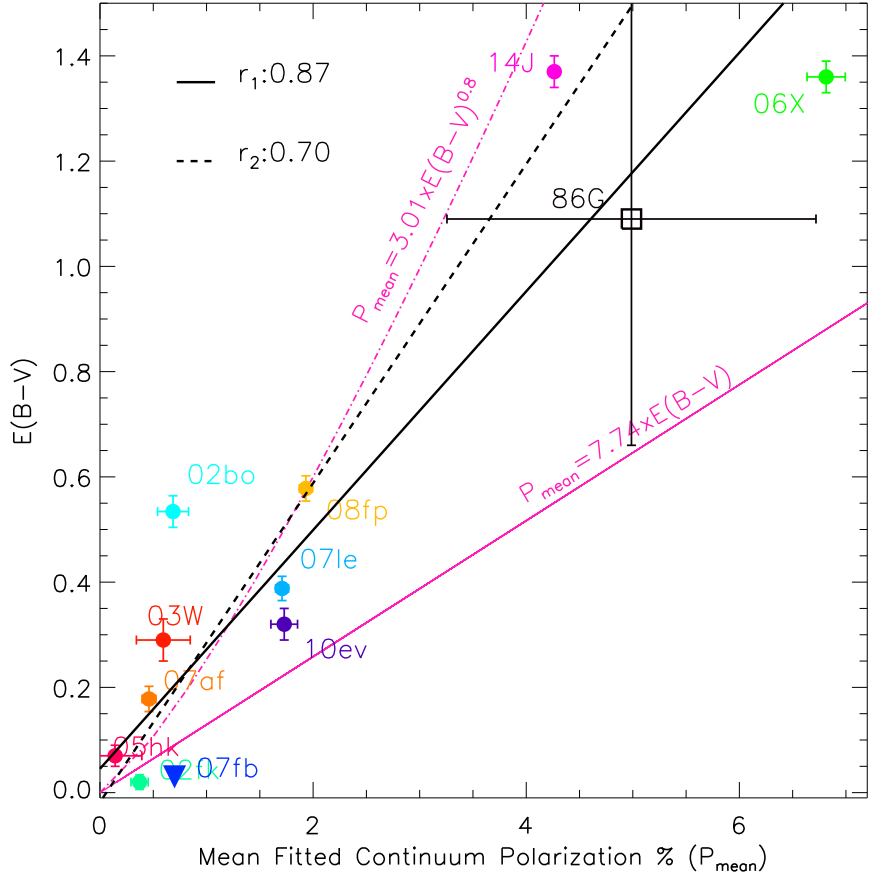}
\caption{
Correlation between the color excess $E(B-V)$ and $P_{\rm{mean}}$.
As in previous figures the black solid line shows the linear fit computed for the entire sample
and the black dashed line the linear fit when SN~2006X is excluded.
The correlation coefficients are r$_{1}$ and r$_{2}$ respectively.
The magenta solid line shows the upper limit of polarization for a given
color excess $E(B-V)$ found by \citet{Serkowski1975ApJ...196..261S} for stars in the Galaxy.
The magenta dot-dashed line is the mean relation found by
\citet{Fosalba2002ApJ...564..762F}, also for stars in the Galaxy, given by their Equation~3.
}
\label{fig:E_BVvsPmean}
\end{figure}

\begin{figure}[t!]
\centering
\includegraphics[width=8cm,angle=0]{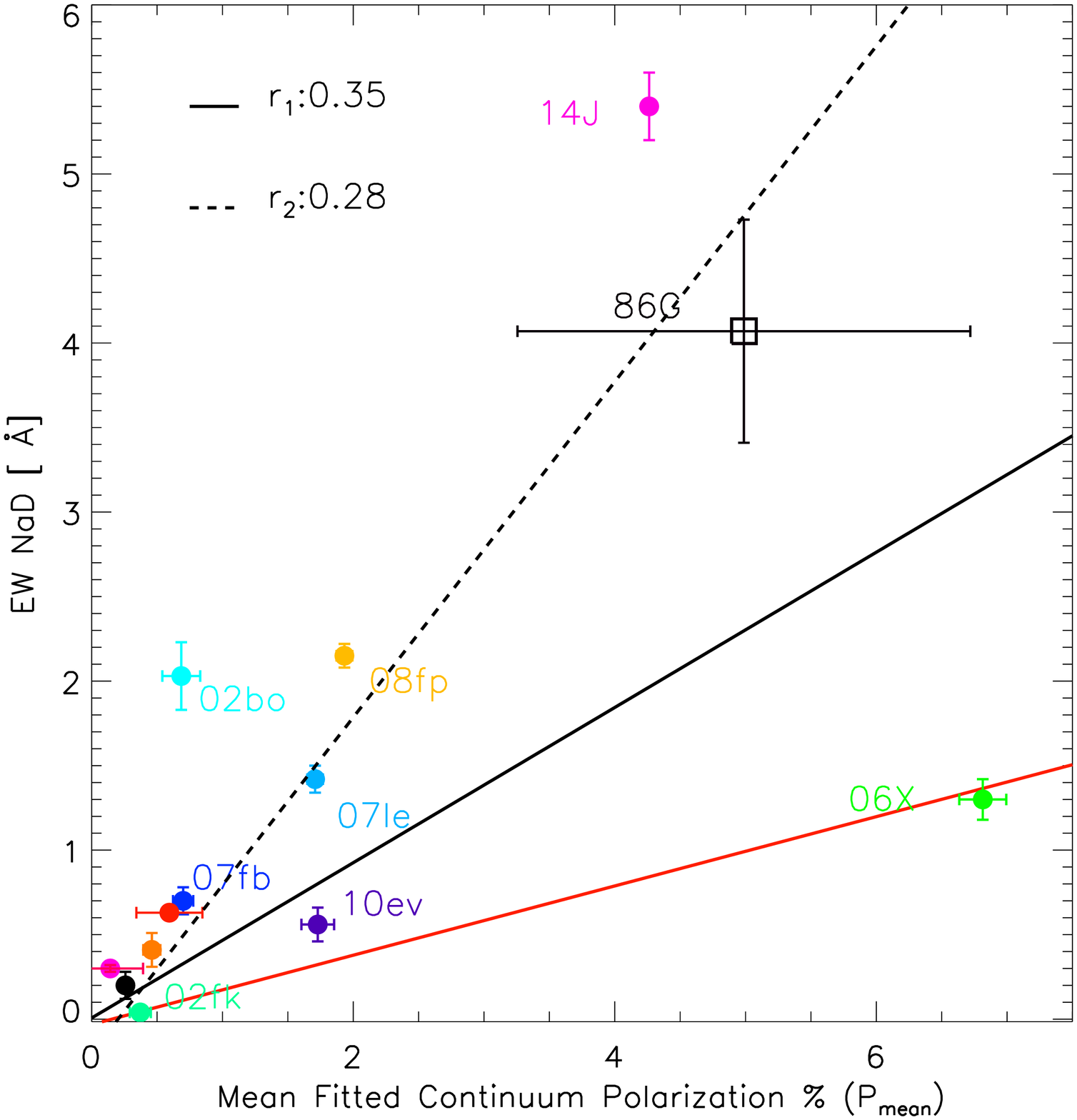}
\includegraphics[width=8cm,angle=0]{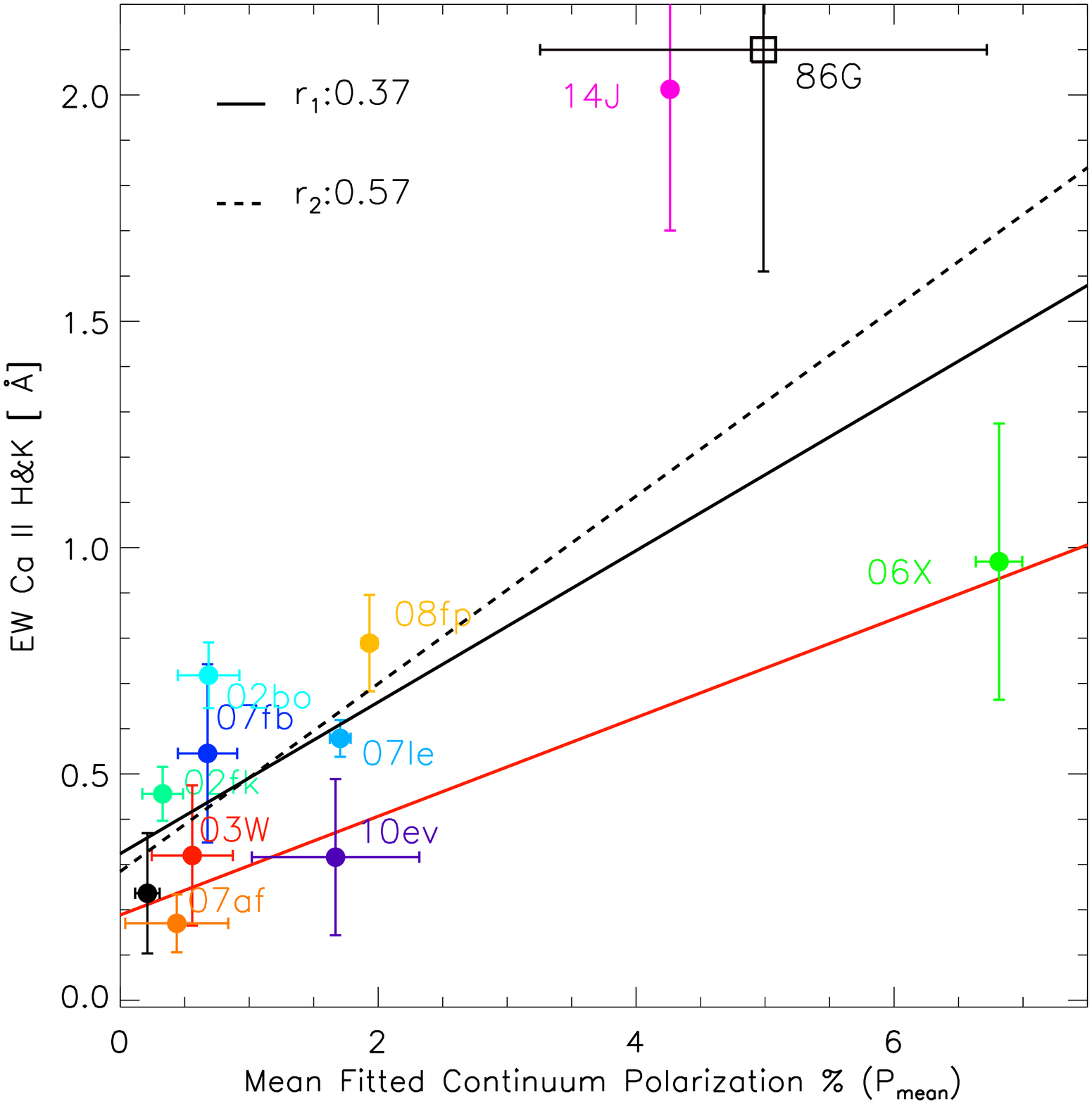}
\includegraphics[width=8cm,angle=0]{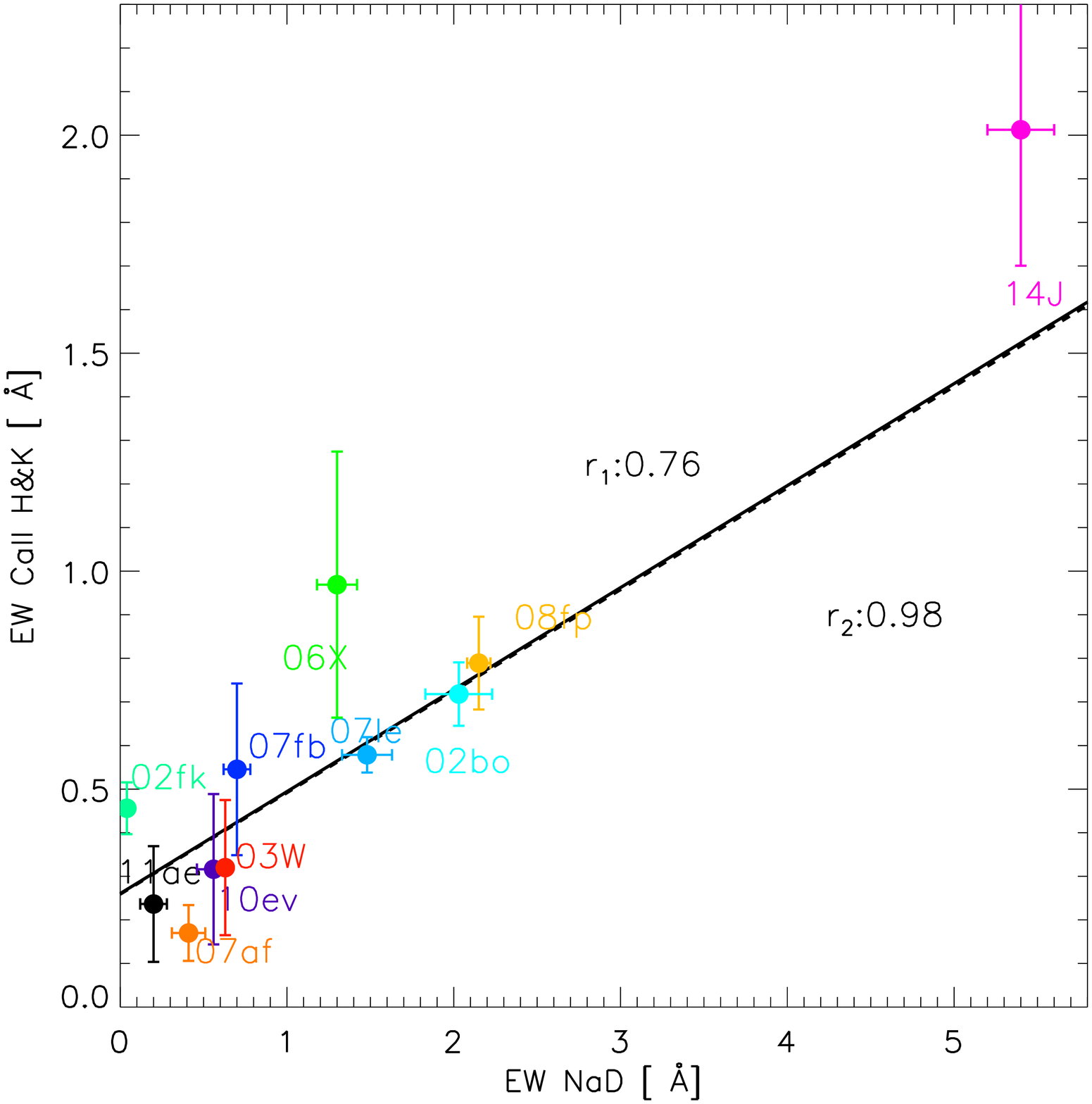}
\caption{
Correlation between the EW of the narrow lines of foreground origin and $P_{\rm{mean}}$.
Na~I~D is shown in the top left panel and Ca~II~H\&K in the top right.
As in previous figures the black solid line shows the linear fit computed for the entire sample
and the black dashed line the linear fit when SN~2006X is excluded.
The solid red lines are approximate upper limits to the observed continuum polarization given the EW of the foreground absorption lines.
The bottom panel shows the correlation between the EW of the Na~I~D and Ca~II~H\&K lines for each SN.
}
\label{fig:fig3}
\end{figure}

\begin{figure}[t!]
\center
\includegraphics[width=8cm,angle=0]{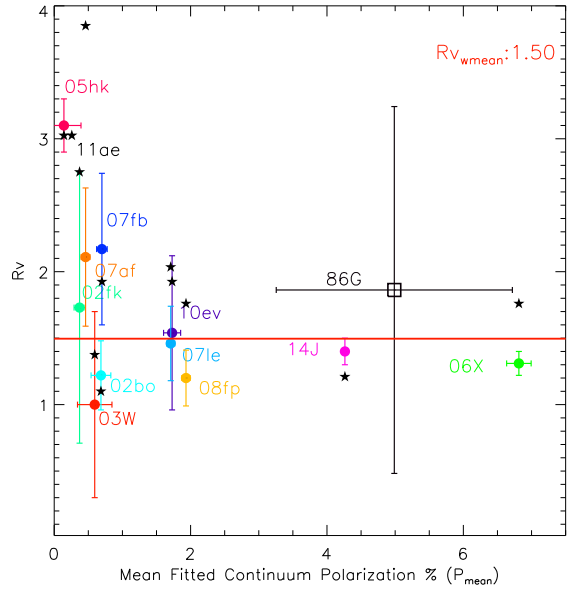}
\caption{
$R_{V}$ values versus the continuum foreground polarization parameter  P$_{\rm{mean}}$.
Black stars correspond to the $R_{V}$ predicted by the relation
$R_{V}$ = 5.5 $\lambda_{\rm{max}}$ found by \citet{Serkowski1975ApJ...196..261S} for stars in the Galaxy.
The red horizontal line is the weighted average of the ten SNe that have a reported uncertainty ($R_{V} = 1.5 \pm 0.06$).
}
\label{fig:fig4}
\end{figure}

The continuum polarization of the sodium sample (see Fig.~\ref{fig:fig1}) is generally well matched by the original ISP
polarization spectrum found by \citet{Serkowski1975ApJ...196..261S}, with constant $K \simeq 1.15$ and values of
$\lambda_{\rm{max}}$ that are short by the standards of the Galaxy. 
Furthermore, the blue continuum polarization is poorly fitted if the $K$ versus $\lambda_{\rm{max}}$ relation found by
\citet{Whittet1992ApJ...386..562W} is used.
For each SNe, the polarization remains approximately constant for the entire period of observation, which may be
as long as several weeks.
This rules out polarization originating in circumstellar LE, which should evolve on shorter time scales
\citep{Wang2005ApJ...635L..33W,Patat2006MNRAS.369.1949P}.
On the other hand, the non-sodium sample (Fig.~\ref{fig:fig1b}) typically shows smaller values of blue polarization and a wavelength
dependence that can be well fitted by a Serkowski profile with values of $\lambda_{\rm{max}}$ similar to those of the
Galaxy, which follow the \citet{Whittet1992ApJ...386..562W} relation.
Additionally, the polarization pseudo-vectors of the sodium sample tend to align with major macroscopic features of the parent galaxies
while those of the non-sodium sample do not.

In the wavelength range 420--580 nm, where polarization intrinsic to the SNe is expected to be
small, the foreground polarization can be reasonably well described by
$P_{\rm{cont}}(\lambda)= a + b\lambda$ (Eq.~\ref{eq:straightfit}).
There is a good correlation between $b$ and $P_{\rm{mean}}$, the mean polarization in the selected blue wavelength range,
which is expected for a Serkowski-type polarization.
This makes it possible to describe the blue continuum polarization with a single observational parameter.

There are good correlations between $P_{\rm{mean}}$, and the color, color excess and $A_V$ of the SNe.
They appear like the usual \textquotedblleft wedge\textquotedblright scatter plots where the dispersion around the mean of the variable taken as dependent
grows together with the variable taken as independent ($P_{\rm{mean}}$ in this case).
A noisier relation also exists between $P_{\rm{mean}}$ and the value of $R_{V}$ estimated from photometry.
The latter is consistent with both extinction and polarization resulting from dust with a
typical size smaller than the characteristic size of polarizing dust in our neighborhood of the Galaxy.
A correlation also exists between $P_{\rm{mean}}$ and the equivalent widths of the foreground Na~I~D and Ca~II~H \& K narrow lines,
which can be also interpreted as upper and lower limits for $P_{\rm{mean}}$, given a column density of gas. 
As is typical of wedge-like scatter plots, these upper and lower limits are quite similar for small polarization, but they diverge to
a large extent at higher polarizations.

The spectropolarimetric observations bring an additional element into the discussion of the location of the
matter that reddens and/or polarizes the SN light.
The polarizing dust has to be aligned with the magnetic field and so has to be in regions that have remained
free from shocks, strong winds, or other large scale hydrodynamic phenomena for times longer than
the time scale for dust alignment.
This rules out a polarizing dust located very close to the SN as in a CSM region.
Still, the fact that the SNe with higher foreground continuum polarization are all blueshifted in the
classification of \citet{Sternberg2011Sci...333..856S}, would suggest that the relevant
polarizing regions could be associated with the SN progenitor, or the star--forming episode that
gave birth to them.
Since we cannot establish the distance between the polarizing dust and the SNe, however, our
observations cannot be used to discern between the major scenarios for Type~Ia SN progenitors and
may be consistent with both of them.

Even though our sample is still small, there are two practical conclusions that follow from this study.
One is that the presence of narrow Na~I~D lines at the redshift of the host galaxies, in Type~Ia SNe,
is an indirect indicator of the presence of dust grains, which have shown to produce smaller $\lambda_{\rm{max}}$ than those of the Milky Way.
Hence its presence is also an indicator of $R_V \lesssim 3$.
In this case, Na~I~D lines at the redshift of the parent galaxy should be taken as an independent prior on
$R_V$ when simultaneously fitting light curves, $K$-corrections, and reddening for Type~Ia SNe, especially for those used in cosmology.
The other conclusion is that the presence of foreground Na~I~D lines at the redshift of the host
impairs the possibility of assuming that the foreground ISP is ``normal'' in the sense of
behaving like the interstellar polarization of the Galaxy.
This should be taken as a warning sign on the prospects of studying the intrinsic continuum polarization of Type~Ia SNe.
Disentangling foreground and intrinsic polarization is, and will continue to be, a risky step.
If the hypothesis that the foreground ISP is well represented by a normal Serkowski law cannot be made,
the only chance of reliably measuring it are late time polarization spectra with very good S/N ratio \citep[like in][]{Patat et2009A&A...508..229P}.
Studies of intrinsic continuum polarization of Type Ia~SNe should be restricted to those that do not show
Na~I~D lines at the redshifts of their hosts, or to those that show sodium, but have an extended time coverage since maximum (more than 60 days).

Our results indicate that the light paths towards extragalactic SNe that show Na~I~D absorption caused by matter in the host galaxies
display a peculiar polarization spectrum. SNe 2006X, 2008fp and 2014J \citep{Patat2015A&A...577A..53P} appear as extreme, highly
polarized, cases of a distribution that extends towards low polarizations as well.
The simplest interpretation is that the lines of sight toward these events preferentially
contain dust with different properties than the dust in the typical line of sight towards stars in the Galaxy.
A scenario of nearby dust clouds accelerated by radiation pressure from the SNe and colliding with more distant ones \citep{Hoang2015arXiv151001822H}
appears to contain many of the elements required to account for the observations, although there are still questions to answer to make it fully convincing.

A better data set to study the properties of these dusty regions should include high--resolution spectroscopy of the 
Na~I~D, K~I, and Ca~II lines, as well as some of the Diffuse Interstellar bands (DIBs), both from the parent galaxies and our Milky Way.
The equivalent width of DIBs is the most accurate tool to probe SN extinction according to \citet{Phillips2013ApJ...779...38P}.
The K~I lines would be especially important since they are less affected by saturation.
The ideal data set should include, as well, a very late spectropolarimetric observation, taken around 50 days after maximum, like the ones we have for
SN~2006X and 2007le.
These data sets would allow an estimate of the temperatures and densities of the dusty gas and help to understand better whether the dust is
related to the SNe or their progenitors, or at least to confirm whether it is under the influence of the SN radiation.
Better data sets also mean, plainly, an improved statistics.
We need more SNe observed in different types of galaxies since the global dust properties are expected to relate to the evolutionary stage of the hosts.
Better observations will allow us, eventually, to bridge the gap between studying the foreground matter in Type~Ia SNe and
connecting the foreground, or parts of it, to a particular SN progenitor scenario.

\acknowledgements Acknowledgments:
We would like to thank Chris Burns and Mark Phillips for their very helpful data calculations and rich discussion.
This paper is based on observations made with ESO Telescopes at the Paranal Observatory under the programs
068.D-0571(A), 069.D-0438(A), 070.D-0111(A), 076.D-0178(A), 079.D-0090(A), 080.D-0108(A), 081.D-0558(A), 085.D-0731(A) and 086.D-0262(A). Also based on observations collected at the German-Spanish Astronomical Center, Calar Alto (Spain).
%

Support for PZ, FF, SG and AC is provided by the Ministry of Economy, Development, and Tourism's
Millennium Science Initiative through grant IC120009, awarded to
The Millennium Institute of Astrophysics, MAS, and by CONICYT through grant Basal CATA PFB 06/09.
The research of JRM is funded through a Royal Society University Research Fellowship.
The research of JCW was supported in part by NSF grant AST-1109801.

\clearpage

\begin{landscape}
\begin{table}[t!] \footnotesize
\caption{\label{tab_obslog} Spectropolarimetry of the Na sample, hosts, observed equivalent widths and polarization}
\begin{tabular}{|c | c | c | c | c | c | c | c | c | c | c | c |}

\hline
SN \,\,\,& date UT \,\,\,& exposure & median & host & galaxy & phase\tablenotemark{b}
\,\,\,&W$_{\lambda}$Na~I\tablenotemark{c} \,\,\,& W$_{\lambda}$Ca~II\tablenotemark{d}\,\,\, & $P_{\rm{mean}}$ &
Na shift\tablenotemark{e}\,\,\, \\
        &                 &       [s]        & airmass   &  & type  &     &  (\AA)                & (\AA)
      &  \%  & \\\hline\hline
06X  & 2006 feb 18.37 & $4\times480 $ & 1.56  & NGC4321 & Sbc    & -1 & 1.30(12) & 0.97(31) & 6.81(18) &B*\\
14J  & 2014 Jan 28.12 & $4\times1200$ & 1.27  &  M82    & I0     & -5 & 5.40(20) & 2.01(31) & 4.26(02) &--  \\
08fp & 2008 Sep 16.35 & $4\times500 $ & 1.54  &ESO428-G14&SAB0\^0(r)pec& -3 & 2.15(07) & 0.79(11) & 1.93(06) &B*\\
07le & 2007 Oct 21.23 & $4\times300 $ & 1.56  & NGC7721 & SA(s)c & -4 & 1.42(08) & 0.58(04) & 1.71(08) &B*\\
10ev & 2010 Jul 06.99 & $4\times360 $ & 1.56  & NGC3244 &SA(rs)cd& -1 & 0.56(10) & 0.32(17) & 1.67(65) &S\\
02bo & 2002 Mar 22.06 & $4\times600 $ & 1.55  & NGC3190 & SA     & -1 & 2.03(20) & 0.72(07) & 0.68(24) &--\\
07fb & 2007 Jul 10.37 & $4\times600 $ & 1.17  & UGC12859& Sbc    & -2 & 0.70(08) & 0.55(20) & 0.68(23) &B\\
03W  & 2003 Feb 03.24 & $4\times1200$ & 1.37  & UGC05234& Sc     & -6 & 0.63(03) & 0.32(16) & 0.56(32) &--\\
07af & 2007 Mar 16.34 & $4\times900 $ & 1.15  & NGC5584 & Scd    & +1 & 0.41(10) & 0.17(06) & 0.44(40) &R\\
02fk & 2002 Oct 01.30 & $4\times720 $ & 1.02  & NGC1309 & SA(s)bc&  0 & 0.04(01) & 0.46(06) & 0.33(16) &--\\
11ae & 2011 Mar 04.19 & $4\times500 $ & 1.06  & UGCA254 &SAB(s)cd& +2 & 0.20(08) & 0.24(13) & 0.21(09) &--\\
05hk & 2005 Nov 09.14 & $4\times900 $ & 1.26  & UGC00272&SAB(s)d & -2 & 0.30(02) & --       & 0.13(37) &--\\

\hline
\end{tabular}
\tablenotetext{b}{Phase (days w.r.t. B maximum light).}
\tablenotetext{c}{Na~I~D equivalent width [\AA].}
\tablenotetext{d}{Ca~II~H\&~K equivalent width [\AA].}
\tablenotetext{e}{Shift of the Na~D lines with respect to a principal Na~D
line component, measured in high resolution spectra from
\citet{Sternberg2011Sci...333..856S}. B, R and S denote blueshifted, redshifted or
single foreground Na~I~D lines, respectively. Symbol * denotes reported
time-variability in some of the Na~I~D lines.}

\end{table}
\end{landscape}

\clearpage

\begin{table}[t!] \footnotesize
\caption{\label{tab_phot} Color, color excess, extinction and $R_{V}$\tablenotemark{1}}
\begin{tabular}{|c | c | c | c | c | c | c | }

\hline
SN &$(B-V)_{Bmax}$& $E(B-V)$   &    $A_V$   &   $R_{V}$  & $A_{V,MW} $ & refs\tablenotemark{2} \\ \hline\hline
06X  & 1.26(05)   & 1.36(03) & 1.88(0.11) & 1.31(0.09) & 0.07 & phi13,bur14,hic09 \\
14J  & 1.25(06)   & 1.37(03) & 1.85(0.11) & 1.40(0.10) & 0.44 & ama14,mar15       \\
08fp & 0.67(01)   & 0.58(02) & 0.71(0.09) & 1.20(0.21) & 0.54 & phi13,bur14       \\
07le & 0.28(04)   & 0.39(02) & 0.54(0.08) & 1.46(0.28) & 0.09 & phi13,bur14,gan10 \\
10ev & 0.36(05)   & 0.32(03) & 0.50(0.18) & 1.54(0.58) & 0.28 & gut11,phi13,burnspc \\
02bo & 0.37(07)   & 0.53(03) & 0.62(0.10) & 1.22(0,26) & 0.07 & phi13,kris04,ben04,hic09,gan10 \\
07fb & -0.06(01)  & $\lesssim$ 0.03 & $\lesssim$ 0.09 & 2.17(0.57)& 0.15 & phi13,burnspc \\
03W  & 0.16(01)   & 0.29(04) & 0.30(19)   & 1.0(0.70)  & 0.13 & burnspc,hic09 \\
07af & 0.08(05)   & 0.15(02) & 0.31(06)   & 2.11(0.51) & 0.10 & phi13,bur14 \\
02fk & -0.12(06)  & 0.02(01) & 0.03(02)   & 1.73(1.02) & 0.11 & burnspc,hic09 \\
11ae & 0.03(02)   & --       & --         & --         & 0.16 & -- \\ 
05hk & -0.03(04)  & 0.07(02) & 0.22(0.06) & 3.1        & 0.06    & mau10,phi07,cho06,hic09,sah08 \\

\hline
\end{tabular}
\tablenotetext{1}{Values corrected for extinction in the Milky Way \protect{\citep{Schlafly2011ApJ...737..103S}.}}
\tablenotetext{2}{
ama14=\citet{Amanullah2014} \\
mar15=\citet{Marion2015ApJ...798...39M} \\
bur14=\citet{Burns2014} \\
burnspc=C.Burns personal communication  \\
phi13=\citet{Phillips2013ApJ...779...38P} \\
gut11=\citet{Gutierrez2011BAAA...54..109G} \\
gan10=\citet{Ganeshalingam2010ApJS..190..418G} \\
mau10=\citet{Maund2010ApJ...722.1162M} \\
hic09=\citet{Hicken2009ApJ...700..331H} \\
sah08=\citet{Sahu2008ApJ...680..580S} \\
phi07=\citet{Phillips2007PASP..119..360P} \\
cho06=\citet{Chornock2006PASP..118..722C} \\
kri04=\citet{Krisciunas2004AJ....128.3034K} \\
ben04=\citet{Benetti2004MNRAS.348..261B} \\
}

\end{table}

\clearpage

\clearpage

\newpage

\newpage

\newpage

\newpage

\newpage

\newpage

\newpage

\clearpage

\clearpage

\clearpage

\clearpage
\newpage

\appendix
\section{Appendix, SN~1986G}
\label{ap:86G} 
\setcounter{figure}{0}  
\counterwithin{figure}{section}

We have included a publicly available spectrum of SN 1986G near maximum light
\footnote{From Weizmann Interactive Supernova data REPository (WISeREP), http://wiserep.weizmann.ac.il/} 
which also has polarimetric data published by \citet{Hough1987MNRAS.227P...1H}.

\begin{figure}[t!]
\center
\includegraphics[width=12cm,angle=0]{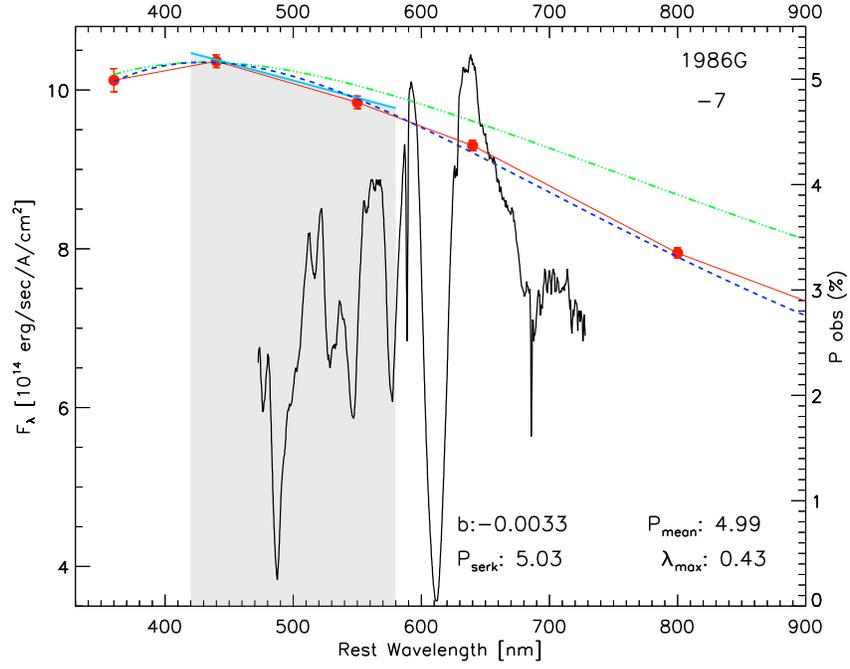}
\caption{Same as Figure~\ref{fig:fig1}, except for imaging polarimetry data of SN~1986G acquired by \citet{Hough1987MNRAS.227P...1H}.
}
\label{fig:86Gpol}
\end{figure}

This is a highly polarized SN Ia with large amounts of sodium in its host, Centaurus A. Although the 
observations are not spectropolarimetric, they make a good addition to our sample.
Different filters used by \citet{Hough1987MNRAS.227P...1H} allow us to model the Serkowski law in the same way as our 
spectropolarimetric sample, but with a much larger error (see Fig.~ \ref{fig:86Gpol}). 
We obtain $\lambda_{\rm{max}}$=0.43$~\mu$m and for the linear fit P$_{\rm{mean}}=4.98 \pm~1.73$ and b=$-0.003\pm~ 0.001$.

Extinction values were obtained from \citet{Phillips2013ApJ...779...38P}, where $A_V=2.03 \pm~0.11$ and (B-V)$_{\rm{Bmax}}=1.08 \pm~0.5$ mag.
The color excess corresponds to an average value E(B-V)=1.09 $\pm$ 0.43 mag from the calculations of 
\citet{Hough1987MNRAS.227P...1H} and \citet{Phillips1987PASP...99..592P,Phillips2013ApJ...779...38P}.

For the equivalent widths of Na~I~D and Ca~II~H\&K \citet{Phillips1987PASP...99..592P} measures 3.6 
and 1.75 \AA{} respectively, while \citet{DOdorico1989A&A...215...21D} find, 4.53 and 2.45 \AA{}. 
We take the average between these two and consider equivalents widths of $4.07\pm0.66$ and $2.10\pm0.49$, respectively.

\begin{figure}[t!]
\center
\includegraphics[width=8cm,angle=0]{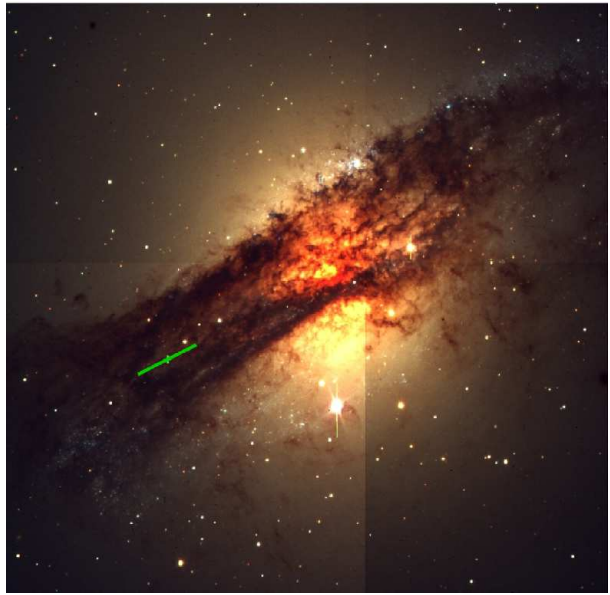}
\caption{Same as Figure~\ref{fig:Host_Na} for SN~1986G in Centaurus A.
}
\label{fig:86Ghost}
\end{figure}

\end{document}